\documentclass{article} %
\usepackage{iclr2023_conference}
\usepackage{times}
\iclrfinalcopy

\usepackage{bm}
\usepackage{amssymb,amsmath,graphicx,bbm,color,booktabs}
\usepackage{amsopn}

\newcommand{\vc}{\bm{c}}

\newcommand{\vu}{\bm{u}}               
\newcommand{\vv}{\bm{v}}               
               
\newcommand{\vx}{\bm{x}}       \newcommand{\vxh}{\hat{\bm{x}}}        
               
\newcommand{\vz}{\bm{z}}       \newcommand{\vzh}{\hat{\bm{z}}}

  \newcommand{\Sc}{\mathcal{S}}

\renewcommand{\eqref}[1]{Eq.~(\ref{#1})}

\usepackage{hyperref}
\usepackage{url}
\usepackage{blindtext}
\usepackage{graphicx}
\usepackage{pifont}
\usepackage{acronym}
\usepackage{multirow}

\newcommand{\pmr}[1]{\scriptsize$\pm$#1}

\usepackage{booktabs}
\usepackage{multirow}

\usepackage{amsmath}
\usepackage{amssymb}
\usepackage{mathtools}
\usepackage{amsthm}

\usepackage{caption}
\usepackage{subcaption}

\renewcommand{\[}{\begin{eqnarray}}
\renewcommand{\]}{\end{eqnarray}}

\newcommand{\audiogen}{\textsc{AudioGen}}
\newcommand{\src}{\textsc{Src}}
\newcommand{\rnd}{\textsc{Rnd}}

\hypersetup{
    colorlinks,
    linkcolor={red!50!black},
    citecolor={blue!50!black},
    urlcolor={blue!80!black}
}

\acrodef{SI-SNR}{Scale-Invariant Signal-to-Noise Ration}
\acrodef{SNR}{Signal-to-Noise Ratio}
\acrodef{CFG}{Classifier Free Guidance}
\acrodef{ALM}{Audio Language Model}
\acrodef{LUT}{Look-Up-Table}
\acrodef{FAD}{Fr{\'e}chet Audio Distance}
\acrodef{FID}{Fr{\'e}chet Inception Distance}
\acrodef{SSL}{Self-Supervised Learning}

\title{AudioGen: Textually Guided Audio Generation}

\author{Felix Kreuk$^{1}$, Gabriel Synnaeve$^{1}$, Adam Polyak$^{1}$, Uriel Singer$^{1}$, Alexandre D\'efossez$^{1}$, \\ \textbf{Jade Copet$^{1}$, Devi Parikh$^{1}$, Yaniv Taigman$^{1}$, Yossi Adi$^{1,2}$} \\\\
$^{1}$FAIR Team, Meta AI\\
$^{2}$The Hebrew University of Jerusalem\\
\texttt{felixkreuk@meta.com}
}

\begin{document}

\maketitle

\begin{abstract}
We tackle the problem of generating audio samples conditioned on descriptive text captions. In this work, we propose \audiogen, an auto-regressive generative model that generates audio samples conditioned on text inputs. \audiogen~~operates on a learnt discrete audio representation. The task of text-to-audio generation poses multiple challenges. Due to the way audio travels through a medium, differentiating ``objects'' can be a difficult task (e.g., separating multiple people simultaneously speaking). This is further complicated by real-world recording conditions (e.g., background noise, reverberation, etc.). Scarce text annotations impose another constraint, limiting the ability to scale models. Finally, modeling high-fidelity audio requires encoding audio at high sampling rate, leading to extremely long sequences. To alleviate the aforementioned challenges we propose an augmentation technique that mixes different audio samples, driving the model to internally learn to separate multiple sources. We curated 10 datasets containing different types of audio and text annotations to handle the scarcity of text-audio data points. For faster inference, we explore the use of multi-stream modeling, allowing the use of shorter sequences while maintaining a similar bitrate and perceptual quality. We apply classifier-free guidance to improve adherence to text. Comparing to the evaluated baselines, \audiogen~outperforms over both objective and subjective metrics. Finally, we explore the ability of the proposed method to generate audio continuation conditionally and unconditionally. Samples: \url{https://felixkreuk.github.io/audiogen}.
\end{abstract}

\vspace{-0.1cm}
\section{Introduction}
\label{sec:intro}
\vspace{-0.1cm}
Neural generative models have challenged the way we create digital content. 
From generating high-quality images~\citep{karras2019style, park2019semantic} 
and speech~\citep{ren2020fastspeech, oord2016wavenet}, through generating long textual spans~\citep{brown2020language, zhang2022opt}, to the recently proposed text prompted image generation~\citep{ramesh2022hierarchical, Rombach_2022_CVPR}, these models have shown impressive results. 
This begs the question \emph{what would be the audio equivalent to textually guided generative models?} 
From generating soundscapes to music or speech, a solution to this problem that is high fidelity, controllable, and diverse in its outputs, would be a useful addition to the modern toolbox of creators of movies, video games, and any virtual environments. 

While image generation and audio generation have a lot in common, there are a few key differences. Audio is intrinsically a one dimensional signal and thus has less degrees of freedom to differentiate overlapping ``objects''~\citep{capon1969high, frost1972algorithm}. Real-world audio inherently has reverberations, which makes the task of differentiating objects from the surrounding environment even harder. Moreover, psychoacoustic and psychovisual properties differ, for instance hearing ``resolution'' (equal-loudness) is U-shaped in frequencies with a dip at 4kHz and bump at 8kHz \citep{suzuki2003precise}. Last but not least, the availability of audio data with textual descriptions is orders of magnitude below that of text-image paired data. This makes generating unseen audio compositions a hard task (e.g. generating an audio equivalent of an image of ``an astronaut riding a horse in space'').

In this work, we tackle the problem of generating audio samples conditioned on descriptive text captions. We additionally extend the proposed method to conditional and unconditional audio continuation. Here, we generate ``a dog barks while somebody plays the trumpet in a busy street''. In the above prompt, the model must generate three categories of acoustic content, with varying degrees of background/foreground, durations, and relative position in the temporal axis, the composition of which is highly unlikely to be present in the training set. Generating such audio is thus a challenge in generalization, acoustic fidelity, production and mastering. 

We propose \audiogen, an autoregressive textually guided audio generation model. \audiogen~~ consists of two main stages. The first encodes raw audio to a discrete sequence of tokens using a neural audio compression model (e.g.~\cite{zeghidour2021soundstream}). This model is trained in an end-to-end fashion to reconstruct the input audio from the compressed representation, with an addition of a perceptual loss in the form of a set of discriminators. Such an audio representation is designed to generate high-fidelity audio samples while still being compact. The second stage, leverages an autoregressive Transformer-decoder language-model that operates on the discrete audio tokens obtained from the first stage while also being conditioned on textual inputs. We represent text using a separate text encoder model pre-trained on a large corpus of text, namely T5~\citep{raffel2020exploring}. The pre-trained text encoder enables the generalization to text concepts that are absent from current text-audio datasets. This is especially important when working with text annotations limited in terms of diversity and descriptiveness.

Compared to the existing text-to-audio work~\citep{yang2022diffsound}, \audiogen~generates samples that obtain better objective and subjective metrics. In particular, \audiogen~creates more natural sounding unseen audio compositions. Lastly, we empirically show how the proposed approach can be extended to audio continuation considering both conditional and unconditional generation.

{\noindent \bf{Our contributions:}}
(i) We propose a state-of-the-art auto-regressive audio generation model conditioned on textual descriptions or audio prompts, as evaluated with objective and subjective (human listeners) scores. Specifically we propose two model variations, one with 285M parameters and another one with 1B parameters;
(ii) We improve text-to-audio generation in two axes. We improve text adherence by applying classifier free guidance on top of the audio language model. We improve compositionality by performing on the fly text and audio mixing; (iii) We show that the proposed approach can be extended to audio continuation conditioned on text and unconditionally; (iv) We explore the trade-off between audio-fidelity and sampling time by utilizing residual vector quantization (for acoustic units) and multi-stream transformers.

\vspace{-0.2cm}
\section{Related Work}
\label{sec:related}
\vspace{-0.1cm}

{\noindent \bf{Speech Representation Learning.}} Studies on unsupervised speech representation learning can be roughly divided into reconstruction and self-supervised learning methods. Auto-encoding is the common approach for signal reconstruction, where speech is first encoded into a low-dimensional latent representation, and then decoded back to speech. Various constraints can be imposed on the encoded space, such as temporal smoothness~\citep{ebbers2017hidden}, discreteness~\citep{van2017neural}, and hierarchy~\citep{hsu2017unsupervised}. \ac{SSL} methods for speech have shown remarkable results for automatic speech recognition~\citep{schneider2019wav2vec, baevski2020wav2vec, wang2021unispeech}, phoneme segmentation~\citep{kreuk2020self}, and audio compression~\citep{zeghidour2021soundstream, polyak2021speech}. \citet{oord2018representation} and \citet{schneider2019wav2vec} suggested training a convolutional neural network to distinguish true future samples from random distractor samples using a Contrastive Predictive Coding (CPC) loss function. \citet{ao2021speecht5} proposed a speech version of the T5 model and showed its efficiency on various speech tasks. Similar to CPC, \citet{baevski2020wav2vec} use an encoder and a predictor, which is trained contrastively to distinguish positive and negative samples. Unlike~\citep{schneider2019wav2vec}, it discretizes and masks segments of the encoder's output. \citet{hsu2021hubert} proposed the HuBERT model which is trained with a masked prediction task similar to BERT~\citep{devlin-etal-2019-bert} but with masked continuous audio signals. \citet{chen2022wavlm} proposed a similar version of HuBERT trained on larger and augmented dataset. More recently, \citet{xu2022masked} proposed a Masked Auto Encoding approach for learning a speech representation and show it efficiency on several audio classification tasks.  

Another line of relevant prior work relates to modeling audio discrete representations. Recent studies suggest quantizing \ac{SSL} representations using k-means and later perform language modeling~\citep{lakhotia2021generative, kharitonov2022textless, borsos2022audiolm}, multi-stream processing~\citep{kharitonov2021text}, speech emotion conversion~\citep{kreuk2021textless}, spoken dialuge~\citep{nguyen2022generative}, and speech-to-speech translation~\citep{lee2021direct, lee-etal-2022-textless, popuri2022enhanced}.

{\noindent \bf{Text-to-Image}} has seen great advances recently. DALL-E~\citep{dalle} first transforms the patches of an image to discrete codes using a pre-trained VQ-VAE. During training, codes representing image patches were appended to codes representing text. Then, a Transformer-decoder model was trained to model these codes in an autoregressive fashion, while \citet{gafni2022make} suggested a similar approach and incorporated segmentation maps for increased controllability. In the Parti model~\citep{parti} the authors suggested modeling the task of text-to-image as a sequence-to-sequence problem using Transformers in an encoder-decoder architecture.   

More recently, diffusion model have gained increased popularity \citep{nichol2021glide,ramesh2022hierarchical,imagen, Rombach_2022_CVPR}. DALLE-2 \citep{ramesh2022hierarchical} used a diffusion model to predict the CLIP visual features given the CLIP text encoding (prior), and another diffusion model to predict the image pixels given the predicted CLIP visual features (decoder). The predicted image is upsampled to a higher resolution using a cascade of super-resolution models. Imagen \citep{imagen} employed a similar approach but omitted the prior component in favor of using pre-trained text-encoders such as T5 \citep{raffel2020exploring}.

{\noindent \bf{Text-to-Audio.}} The most relevant to our work is the one proposed by \cite{yang2022diffsound}, in which the authors proposed DiffSound, a text-to-audio model based on a diffusion process that operates on audio discrete codes. The audio codes were obtained from a VQ-VAE~\citep{NIPS2017_7a98af17} based model trained over mel-spectrogram. To further improve model performance, \citet{yang2022diffsound} suggested pre-training the diffusion model using labeled tags with a random input masking. They additionally explore the usage of an auto-regressive Transformer decoder model, however found it to be inferior to the diffusion based model. 

The proposed method differentiate from DiffSound in the following: (i) our audio representation is being learned directly from the raw-waveform; (ii) we create new audio compositions using data augmentation allowing the model to generate audio from complex text captions; (iii) we apply and study the effect of classifier free guidance under the auto-regressive setting; (iv) in contrast to~\cite{yang2022diffsound}, we empirically demonstrate that text-conditioned auto-regressive models can generate high-quality audio samples.

\vspace{-0.2cm}
\section{Method}
\begin{figure}[t!]
  \centering
  \includegraphics[width=0.85\linewidth]{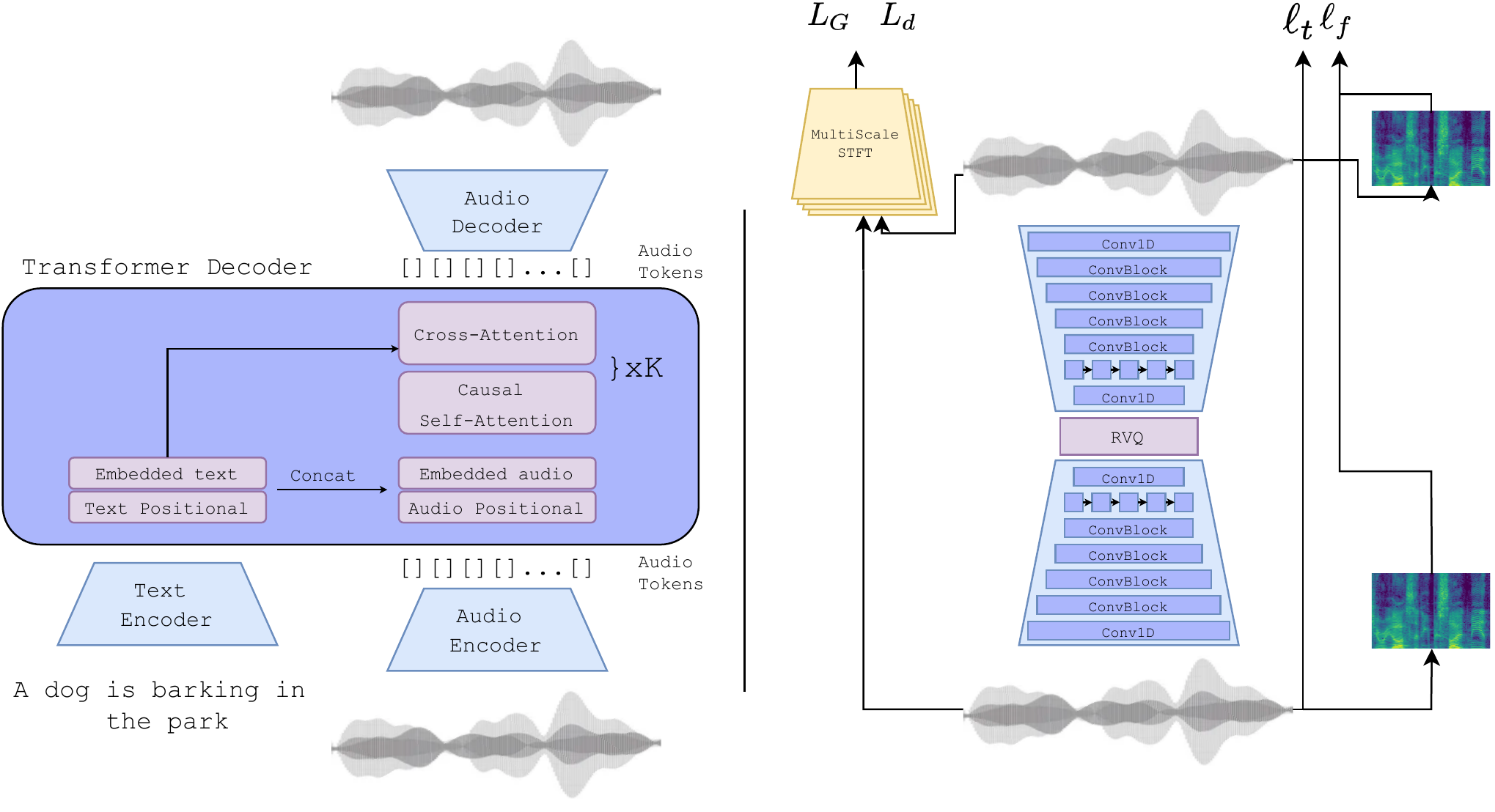}
  \caption{A general overview of the \audiogen system. Left: the audio representation model. Right: the audio language model. Both text and audio embeddings are concatenated over the time dimension and fed in K causal self-attention and cross-attention blocks with the embedded text.}
  \label{fig:arch}
\end{figure}

\vspace{-0.1cm}
\label{sec:method}
The proposed method, \audiogen, is based on two main steps: (i) learning discrete representation of the raw audio using an auto-encoding method; (ii) training a Transformer language model over the learnt codes obtained from the audio encoder, conditioned on textual features. Then, during inference time, we sample from the language model to generate a new set of audio tokens given text features. These tokens can later be decoded into the waveform domain using the decoder component from step (i). A visual description of the proposed method can be seen on Figure~\ref{fig:arch}.

\subsection{Audio Representation}
An audio signal of duration $d$ can be represented by a sequence $\vx\in [-1, 1]^{C_\mathrm{a}\times T}$ with $C_\mathrm{a}$ the number of audio channels, $T = d\cdot f_\mathrm{sr}$  the number of audio samples at a
given sample rate $f_\mathrm{sr}$. In this work we set $f_\mathrm{sr}=16$kHz.
The audio representation model is composed of three components: (i) an encoder network $E$ which gets as input an audio segment and outputs a latent representation $\vz$; (ii) a quantization layer $Q$ produces a compressed representation $\vz_q$, using a Vector Quantization~\citep{vasuki2006review} layer; (iii) a decoder network $G$ reconstructs the time-domain signal, $\vxh$, from the compressed latent representation $\vz_q$. The whole system is trained end-to-end to minimize a reconstruction loss applied over both time and frequency domain, together with a perceptual loss in the form of several discriminators operating at different temporal resolutions. Using a pre-trained model, we can leverage the encoder and quantizer components as a discrete feature extractor (i.e., $Q \circ E$) and $G$ to decode the representation to the time-domain signal. For $Q$, we use a single codebook with 2048 codes, where each code is a 128 dimensional vector. A visual description of the proposed method can be seen in Figure~\ref{fig:arch} (right). 

{\noindent \bf{Architecture.}} We follow a similar auto-encoder model architecture as in~\cite{zeghidour2021soundstream, li2021real}. The encoder model $E$ consists of a 1D convolution with $C$ channels followed by $B$ convolutional blocks. Each convolutional block is composed of a single residual unit followed by a down-sampling layer consisting of a strided convolution, with a kernel size $K$ of twice the stride $S$. The residual unit contains two convolutions and a skip-connection. The number of channels is doubled whenever down-sampling occurs. The convolutional blocks are followed by a two-layer LSTM for sequence modeling and a final 1D convolution layer with a kernel size of 7 and $D$ output channels. We use $C$ = 32, $B$ = 4 and (2, 2, 2, 4) as strides. We use ELU as a non-linear activation function~\citep{clevert2015fast} and a LayerNorm~\citep{ba2016layer}. The decoder mirrors the encoder, using transposed convolutions instead of strided convolutions, and with the strides in reverse order as in the encoder, outputting the final audio.

{\noindent \bf{Training Objective.}} We optimize a GAN based training objective similar to~\citep{kong2020hifi, zeghidour2021soundstream} of jointly minimizing a combination of reconstruction losses and adversarial losses.
Specifically, we minimize the L1 distance between the target and reconstructed audio over the time domain, i.e. $\ell_{t}(\vx, \vxh) = \|\vx - \vxh\|_1$. For the frequency domain loss, we use a linear combination between the L1 and L2 losses over the mel-spectrogram using several time scales~\citep{yamamoto2020parallel, gritsenko2020spectral}. 
Formally, 
\begin{equation}
    \ell_{f}(\vx, \vxh) = \frac{1}{|\alpha|\cdot |s|}\sum_{\alpha_i \in \alpha} \sum_{i \in e} \|\Sc_i(\vx) - \Sc_i(\vxh)\|_1 + \alpha_i \|\Sc_i(\vx) - \Sc_i(\vxh)\|_2,
    \label{eq:msspec}
\end{equation}
where $\Sc_i$ is a 64-bins mel-spectrogram using a normalized STFT with window size of $2^i$ and hop length of $2^i/4$, $e = {5, \ldots, 11}$ is the set of scales, and $\alpha$ represents the set of scalar coefficients balancing between the L1 and L2 terms. Unlike \cite{gritsenko2020spectral}, we set $\alpha_i = 1$.

To further improve the quality of the generated samples, we additionally optimize a multi-scale STFT-based (MS-STFT) discriminator. Multi-scale discriminators are popular for capturing different structures in audio signals~\citep{melgan, kong2020hifi, you2021gan}. The MS-STFT discriminator is based on identically structured networks operating on multi-scaled complex-valued STFT where its real and imaginary parts are concatenated. Each sub-network is composed of a 2D convolutional layer (using kernel size 3 x 8 with 32 channels), followed by 2D convolutions with increasing dilation rates in the time dimension of 1, 2 and 4, and a stride of 2 over the frequency axis. A final 2D convolution with kernel size 3 x 3 and stride (1, 1) provide the final prediction. We use 5 different scales with STFT window lengths of [2048, 1024, 512, 256, 128]. 
The adversarial loss for the generator is constructed as follows, $\ell_{g}(\vxh) = \frac{1}{K} \sum_{k} \max(0, 1 - D_k(\vxh)))$, where $K$ is the number of discriminator networks. Similarly to previous work on neural vocoders~\citep{melgan, kong2020hifi, you2021gan}, we additionally include a feature matching loss for the generator. Formally,
\begin{equation}
    \ell_{feat}(\vx, \vxh) = \frac{1}{KL} \sum_{k=1}^{K}\sum_{l=1}^{L} \| D_{k}^{l}(\vx) - D_{k}^{l}(\vxh) \|_1,
    \label{eq:feat}
\end{equation}
where $(D_k)$ are the discriminators, and $L$ is the number of layers in discriminators. 

Overall the discriminators are trained to minimize the following: $L_{d}(\vx, \vxh) = \frac{1}{K} \sum_{k=1}^{K} \max(0, 1 - D_k(\vx)) + \max(0, 1 + D_k(\vxh))$, where $K$ is the number of discriminators, while the generator is trained to minimize the following: $L_{G} = \lambda_t\cdot\ell_{t}(\vx, \vxh) + \lambda_f\cdot\ell_{f}(\vx, \vxh) + \lambda_g\cdot\ell_{g}(\vxh) + \lambda_{feat}\cdot\ell_{feat}(\vx, \vxh)$.

\subsection{Audio Language Modeling}
Recall, in this work our goal is to generate audio conditioned on text. Specifically, given a textual input $\vc$ the \ac{ALM} component outputs a sequence of audio tokens $\vzh_q$, which can be later decoded into raw audio using $G$. 

Consider a text encoder $F$ which maps a raw text input into a semantic dense representation, $F(\vc) = \vu$. Then, a \ac{LUT} embeds the audio tokens, $\vzh_q$, into a continuous space, $\textrm{LUT}(\vzh_q) = \vv$. We then concatenate both $\vu$ and $\vv$ to create $Z = \vu_1, \dots, \vu_{T_u}, \vv_1, \dots, \vv_{T_v}$, where $T_u$ and $T_v$ are the length of the text representation and audio representation respectively. 

Using the above representation, we train a Transformer-decoder language-model parameterized by $\theta$ using the cross-entropy loss function:
\begin{equation}
    L_{\textrm{LM}} = - \sum_{i=1}^{N} \sum_{j=1}^{T_v} \log p_{\theta}(\vv^i_j | \vu^1_{1}, \dots, \vu^i_{T_u}, \vv^1_{1}, \dots, \vv^i_{j-1}).
\end{equation}

The text representation is obtained using a pre-trained T5 text-encoder ~\citep{raffel2020exploring}. We additionally experimented with learning text embeddings using a \ac{LUT}. Although it produces comparable results to the T5 model, it limits our ability to generalize to unseen words during training, hence we did not pursue this direction. The transformer-decoder language-model is implemented using a GPT2-like architecture~\citep{radford2019language}. To achieve a better text adherence we add cross-attention between audio and text to each attention block of the transformer. See a visual description on Figure~\ref{fig:arch} (left). The entire system may be alternatively viewed as an encoder-decoder model, where the encoder (T5) is pre-trained and fixed throughout training.

{\noindent \bf{Classifier Free-Guidance.}} It was recently shown by \cite{ho2021classifier, nichol2021glide} that using the \ac{CFG} method is an effective mechanism for controlling the trade-off between sample quality and diversity. Although the \ac{CFG} method was originally proposed for the score function estimates of diffusion models, in this work we apply it to auto-regressive models. During training we optimize the Transformer-LM conditionally and unconditionally. In practice, we randomly omit the text conditioning in 10\% of training samples. At inference time we sample from a distribution obtained by a linear combination of the conditional and unconditional probabilities. Formally we sample from,
\begin{equation}
    \gamma \log p_{\theta}(\vv^i_j | \vu^1_{1}, \dots, \vu^i_{T_u}, \vv^1_{1}, \dots, \vv^i_{j-1}) + (1 - \gamma) \log p_{\theta}(\vv^i_j | \vv^1_{1}, \dots, \vv^i_{j-1}), 
\end{equation}
where $\gamma$ is the guidance scale.

{\noindent \bf{Multi-stream audio inputs.}} In order to generate high-quality audio samples we down-sample the raw audio by a factor of 32, which corresponds to 2 ms for each audio token. This requires us to operate over extremely long sequences as each second of audio is represented by 500 tokens. Modeling such long sequences is a notoriously difficult problem \citep{rae2019compressive, zaheer2020big, beltagy2020longformer}. To alleviate this problem, we propose a \emph{Multi-Stream} representation and modeling paradigm. It was shown by~\cite{kharitonov2021text} that transformers are capable of modeling multiple streams simultaneously. 

Consider a sequence of length $T_{v}$, we can learn a representation of length $T_{v}/2$ using two parallel streams of approximately the same bit-rate. This approach can be generalized to $k$ streams, where each stream is of length $T_{v}/k$ and each codebook is of size $2048/k$. Such representation can be obtained by generalizing $Q$ from a single code book Vector-Quantization to a Residual Vector Quantization module as done by~\citet{zeghidour2021soundstream}. At time $t$, the network is fed with $k$ discrete codes, which then embedded using $k$ embedding layers. The final embedding at time $t$ is the mean of these $k$ embeddings. We adapt the network to output $k$ codes using $k$ LM prediction heads. The prediction heads operate independently of each other, we explored conditioning stream $i$ on stream $i-1$ but did not observe any performance gains.  
\section{Experiments}
\label{sec:exp}
In this section we start by providing a detailed description of the experimental setup. Next, we present the main results for audio generation, and we conclude this section with an ablation study.

\subsection{Experimental Setup}
\label{sec:setup}
{\noindent \bf{Dataset.}} We use a set of several datasets: AudioSet~\citep{gemmeke2017audio}, BBC sound effects~\footnote{\url{https://sound-effects.bbcrewind.co.uk/}}, AudioCaps~\citep{kim2019audiocaps}, Clotho v2~\citep{drossos2020clotho}, VGG-Sound~\citep{chen2020vggsound}, FSD50K~\citep{fonseca2021fsd50k}, Free To Use Sounds~\footnote{\url{https://www.freetousesounds.com/all-in-one-bundle/}}, Sonniss Game Effects~\footnote{\url{https://sonniss.com/gameaudiogdc}}, WeSoundEffects~\footnote{\url{https://wesoundeffects.com/we-sound-effects-bundle-2020/}}, Paramount Motion - Odeon Cinematic Sound Effects~\footnote{\url{https://www.paramountmotion.com/odeon-sound-effects}}. All audio files were sampled at 16kHz.

For textual descriptions we use two types of annotations. The first one is multi-label annotations, available for the datasets: AudioSet, VGG-Sound, FSD50K, Sinniss Game Effects, WeSoundEffects, Paramount Motion - Odeon Cinematic Sound Effects. We form pseudo-sentences by concatenating lists of tags available per audio samples (e.g., ``dog, bark, park" is transformed to "dog bark park"). The second type of annotation is natural language captions available for the datasets: AudioCaps, Clotho v2, Free To Use Sounds, and BBC Sound Effects. A more elaborate description of the used datasets can be found in Appendix~\ref{tab:datasets}.
We apply a pre-processing step to better match the class-label annotation distribution. Specifically, we remove stop words and numbers, finally we lemmatize the remaining words (e.g. ``a dog is barking at the park" is transformed to "dog bark park") using the WordNet lemmatizer in NLTK~\citep{bird2009natural}. As speech is the dominant class in the data, we filter all samples where the tag or caption contains the word ``speech'' to generate a more balanced dataset. Overall we are left with $\sim$4k hours for training data.

{\noindent \bf{Data Augmentations.}} One of the most impressive capabilities of recently proposed generative models~\citep{ramesh2022hierarchical, imagen, gafni2022make} is their ability to create unseen object compositions (e.g., ``An astronaut riding a horse in space''). To achieve similar capabilities with regards to audio generation we propose an augmentation method that fuses pairs of audio samples and their respective text captions, thus creating new concept compositions during training. Formally, given two audio samples $\vx_1, \vx_2$ and their respective text captions $\vc_1, \vc_2$, we first randomly draw an temporal offset to merge the two audio samples. Next, we draw a random \ac{SNR} in the interval $[-5,5]$, and finally we mix the audio samples and concatenate the text captions $\vc_1, \vc_2$.

{\noindent \bf{Evaluation Methods.}}
We evaluate all models and baselines using both objective and subjective metrics. For the objective functions we compute the \ac{FAD}~\citep{Kilgour2019FrchetAD} over both real and generated samples. \ac{FAD}, adapted from the \ac{FID} to the audio domain, is a reference-free evaluation metric that closely correlates with human perception. Similarly to~\citet{yang2022diffsound} we additionally compute the KL-Divergence between the output of a state-of-the-art audio classification model~\citep{koutini2021efficient} while feeding it both the original samples and the generated audio.
The FAD was shown to correlate well with human perception in terms of audio quality. On the other hand, the KL measure is computed using the label distribution produced by a pre-trained classifier. Hence, it reflects more on the broader audio concepts occurring in the recording. As a result, the two metrics are complementary. 

For subjective methods, we follow the a similar setting to~\citep{yang2022diffsound}. We ask human raters to evaluate two main aspects of the audio signal (i) overall quality (OVL), and (ii) relevance to the text input. We follow the MUSHRA protocol~\citep{series2014method}, using both a hidden reference and a low anchor. For the overall quality test raters were asked to rate the perceptual quality of the provided samples in a range between 1 to 100. For the text relevance test, raters were asked to rate the match between audio and text on a scale between 1 to 100. Raters were recruited using the Amazon Mechanical Turk platform. We evaluate 100 randomly sampled files from the AudioCaps test set, where each sample was evaluated by at least 5 raters. We verified that the majority of the samples (85\%) contain at least two audio concepts (e.g., ``a dog barks while a bird chirps''). We use the CrowdMOS package\footnote{\url{http://www.crowdmos.org/download/}} to filter noisy annotations and outliers. We remove annotators who did not listen to the full recordings, annotators who rate the reference recordings less then 85, and the rest of the recommended recipes from the CrowdMOS~\citep{ribeiro2011crowdmos}. Participants in this study were paid at least the American minimum wage.

{\noindent \bf{Hyper-parameters.}} We trained two sets of \ac{ALM}s, one with 285M parameters (base) and the other with 1B parameters (large). In the smaller model we use a hidden-size of 768, 24 layers and 16 attention-heads, while for the large variant we use a hidden size 1280, 36 layers and 20 attention-heads. We use the Adam optimizer with a batch size of 256, a learning rate of 5e-4 and 3k steps of warm-up followed by inverse-square root decay. The small model was trained on 64 A100 GPUs for 200k steps ($\sim$5 days) and the large model was trained on 128 A100 GPUs for 200k steps ($\sim$1 week). For the small model we use T5-base and for the large model we use T5-large. For sampling, we employ top-$p$~\citep{holtzman2019curious} sampling with $p=0.25$. For the \ac{CFG} we use a $\gamma=3.0$. 

\begin{table}[t!]
\centering \small
\caption{\label{tab:main_res} Results are reported for DiffSound together with several versions of \audiogen. For DiffSound data augmentation, we follow the authors suggested mask-based text generation (MBTG) strategy. For subjective tests we report overall quality (OVL), and text relevenace (REL.) together with 95\% Confidence Interval. For the objective metrics we report FAD and KL. }
\resizebox{\columnwidth}{!}{%
\begin{tabular}{l|ccc|cc|cc}
\toprule
& & & & \multicolumn{2}{c}{\textsc{Subjective}} & \multicolumn{2}{c}{\textsc{Objective}} \\
 & \#params & \textsc{Aug.} & \textsc{Text-cond.} & \textsc{OVL}$\uparrow$ & \textsc{Rel.}$\uparrow$ & \textsc{FAD}$\downarrow$ & \textsc{KL}$\downarrow$  \\
\midrule
Reference & - & - & - & 92.08\pmr{1.16} & 92.97\pmr{0.85} & - & - \\
\midrule
DiffSound  & 400M & MBTG & CLIP &  65.68\pmr{1.58}  & 55.91\pmr{1.75} & 7.39 & 2.57 \\
\audiogen-base & 285M  & - & T5-base &  70.85\pmr{1.06}  & 63.23\pmr{1.65} & 2.84 & 2.14\\
\audiogen-base & 285M & Mix & T5-base & \bf{71.68\pmr{1.89}}  & 66.01\pmr{1.79} & 3.13 & 2.09\\
\audiogen-large & 1B & Mix & T5-large & \bf{71.85\pmr{1.07}}  & \bf{68.73\pmr{1.61}} & \bf{1.82} & \bf{1.69} \\
\bottomrule
\end{tabular}}
\end{table}

\begin{figure}[t!]
\centering
\begin{subfigure}{0.32\textwidth}
    \centering
    \includegraphics[width=\textwidth]{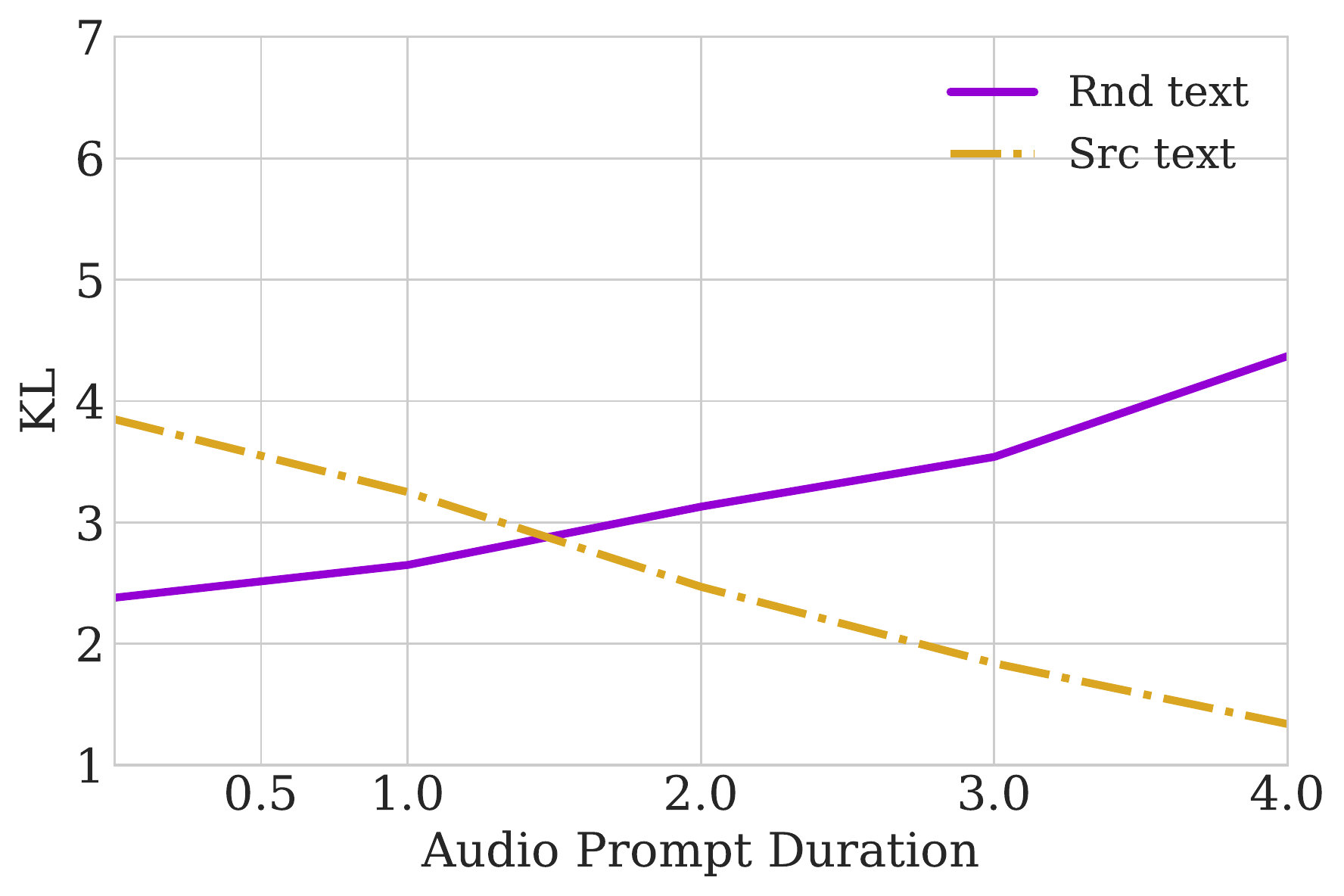}
    \caption{\label{fig:first}}
\end{subfigure}
\begin{subfigure}{0.32\textwidth}
    \centering
    \includegraphics[width=\textwidth]{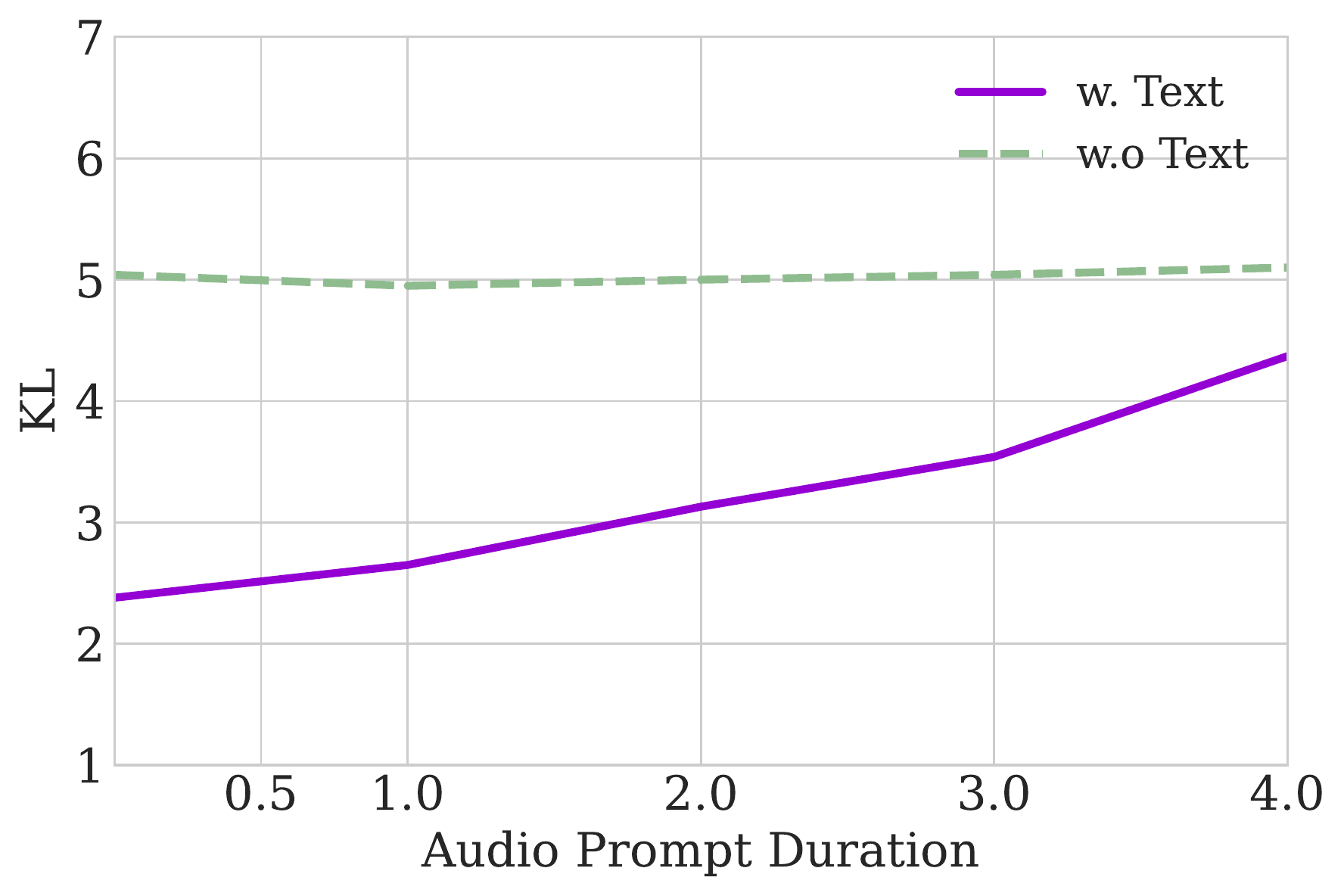}
    \caption{\label{fig:mid}}
\end{subfigure}
\begin{subfigure}{0.32\textwidth}
    \centering
    \includegraphics[width=\textwidth]{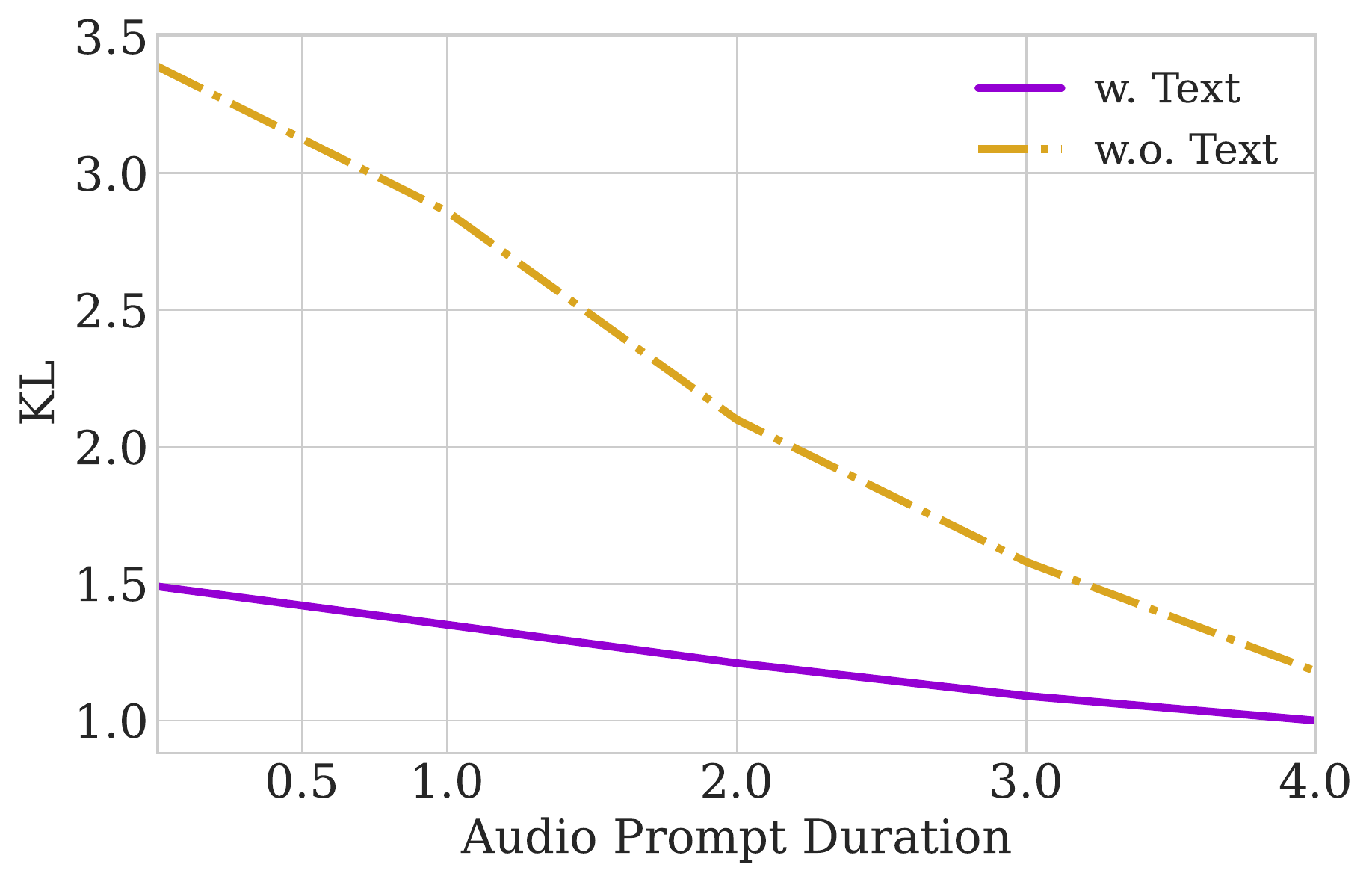}
    \caption{\label{fig:second}}
\end{subfigure}
\caption{\label{fig:audio_cont} Audio continuation results. In (a) we compare model generations against audio corresponding to \src~ and \rnd. In (b) we compare model generations with and without text conditioning against audio corresponding to \rnd~ text. In (c) we compare model generations with and without text conditioning against audio corresponding to \src~ text. In all settings we report KL results. Specific details can be found on Section~\ref{sec:res}.}
\end{figure}

\subsection{Results}
\label{sec:res}

We start by comparing \audiogen~~ to DiffSound. Objective metrics are reported on the AudioCaps test set. We use the official pre-trained model provided by DiffSound authors~\footnote{Pre-trained model can be found under https://github.com/yangdongchao/Text-to-sound-Synthesis}. The model was trained on AudioSet and fine-tuned on AudioCaps.
Results are presented in Table~\ref{tab:main_res}. 

\audiogen-base outperforms DiffSound considering all metrics while being smaller in terms of parameters count. As expected \audiogen-large, significantly outperforms both DiffSound and \audiogen-base. Notice, for \audiogen-base, the model trained without mixing obtained superior \ac{FAD} and comparable OVL to the model trained with mixing augmentations. However, when considering relevance to text, i.e., KL and \textsc{Rel.} the model with mixing augmentations reaches better performance. This is especially noticeable for the \textsc{Rel.} metric as it contains mostly complex compositions (see Table~\ref{tab:main_res}). We additionally compare \audiogen-base to the same dataset setup used by DiffSound (i.e., training on AudioSet and AudioCaps). \audiogen-base reached KL of 2.46 vs. 2.57 for DiffSound and FAD of 4.39 vs. 7.39 for DiffSound. These results suggest that \audiogen is still significantly superior to DiffSound.

Next, we experimented with pre-training the \ac{ALM} component on only audio tokens without conditioning on text (learning an audio prior). 
We did not observe a gain in doing \ac{ALM} pretraining. We hypothesize that is due to the pre-training process taking place on the same labeled data. Samples can be found under the following link: \url{https://felixkreuk.github.io/audiogen}.

{\noindent \bf{Audio Continuation.}} Similarly to~\citep{lakhotia2021generative}, we first encode an audio prompt into a sequence of audio tokens, feed it into the \ac{ALM}, and sample a continuation. Unlike previous work, we can additionally steer the generated audio towards textual captions.
We evaluate our model given audio prompts of lengths [0.5s, 1s, 2s, 3s, 4s] and different combinations of text prompts: (i) no-text; (ii) text corresponding to the audio prompt (\src); (iii) text taken from a randomly sampled audio file (\rnd). For each of the above settings we measure the KL between the generated audio and either the audio corresponding to the source text or the target text. 

In Figure~\ref{fig:audio_cont}(a) we input the model with an audio prompt together with \rnd as conditioning text. We evaluate KL between the output of the classification model using the generated audio against either the audio prompt or the audio corresponding to the \rnd. Results suggest that by using short audio prompts we could better steer the generation towards the conditioning text. In contrast, using long audio prompts leaves less room for textual guidance. Text and audio prompts have roughly the same impact at $\sim$1.5s. In Figure~\ref{fig:audio_cont}(b) we input the model with an audio prompt with and without \rnd text conditioning. We evaluate the KL between the output of the audio classification model using the generated audio against the audio corresponding to the \rnd. Although using longer audio prompts leaves less room for textual guidance, we still observe a gap between generations with and without text, even when using longer audio prompts ($\sim$4s). This suggests that text has a significant effect on the generation process. Lastly, In Figure~\ref{fig:audio_cont}(c) we condition the model on an audio prompt with and without \src text conditioning. We evaluate the KL between the output of the audio classification model using the generated audio against the input audio. Results suggest that when using short audio prompts, text has a major impact on the generated output. However as we feed longer sequences, the audio is sufficient to steer the generation towards the target concept classes with a minimal gain when adding text.

Full results together with FAD scores for all settings together with a visual depiction can be found in Table~\ref{tab:full_audio_cont} on the supplementary \ref{sec:app_res}.

\subsection{Ablation Study.}
\label{sec:abl}

\begin{figure}
\centering
\begin{subfigure}{0.45\textwidth}
    \centering
    \includegraphics[width=\textwidth]{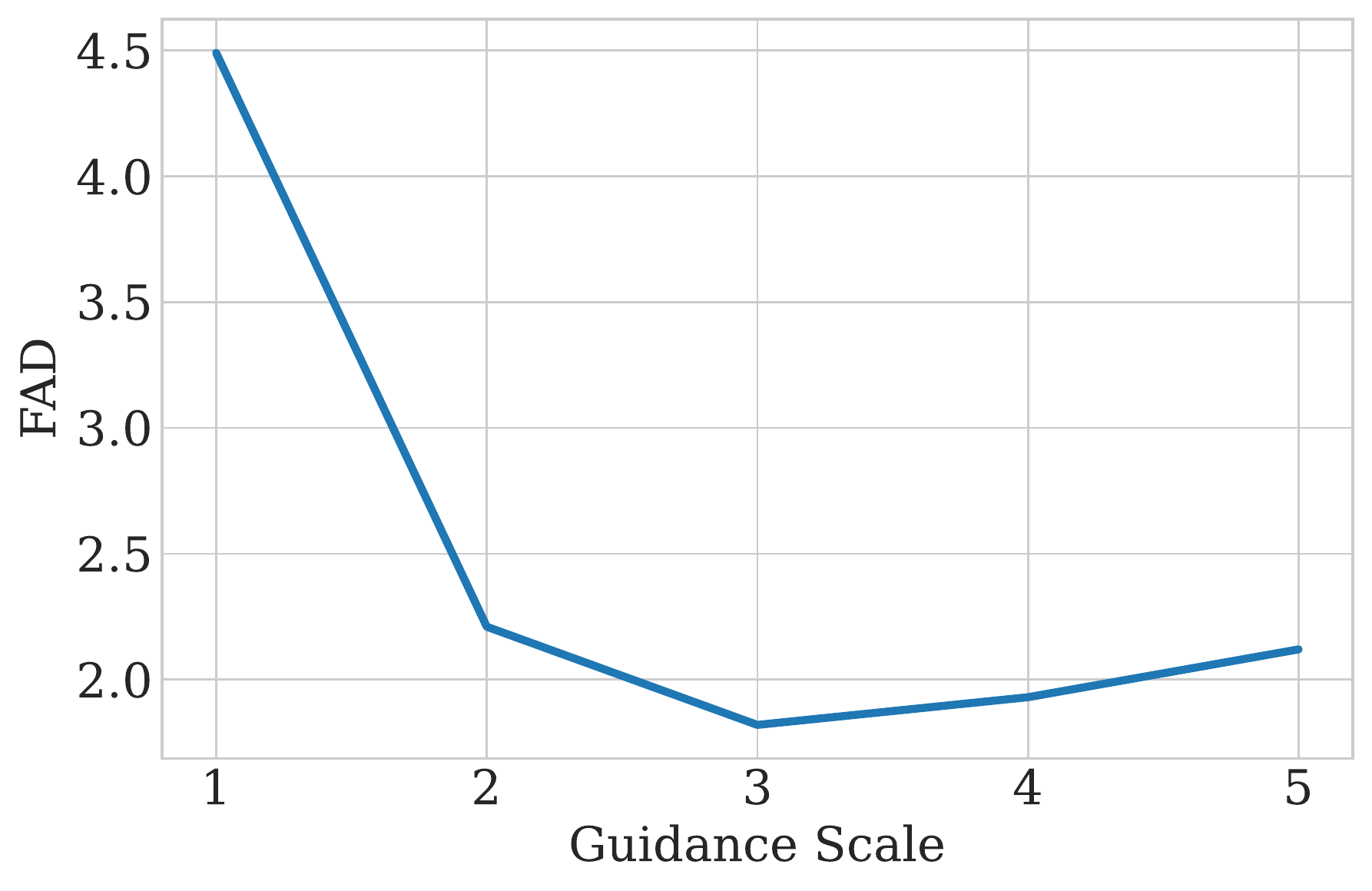}
    \label{fig:first}
\end{subfigure}
\begin{subfigure}{0.46\textwidth}
    \centering
    \includegraphics[width=\textwidth]{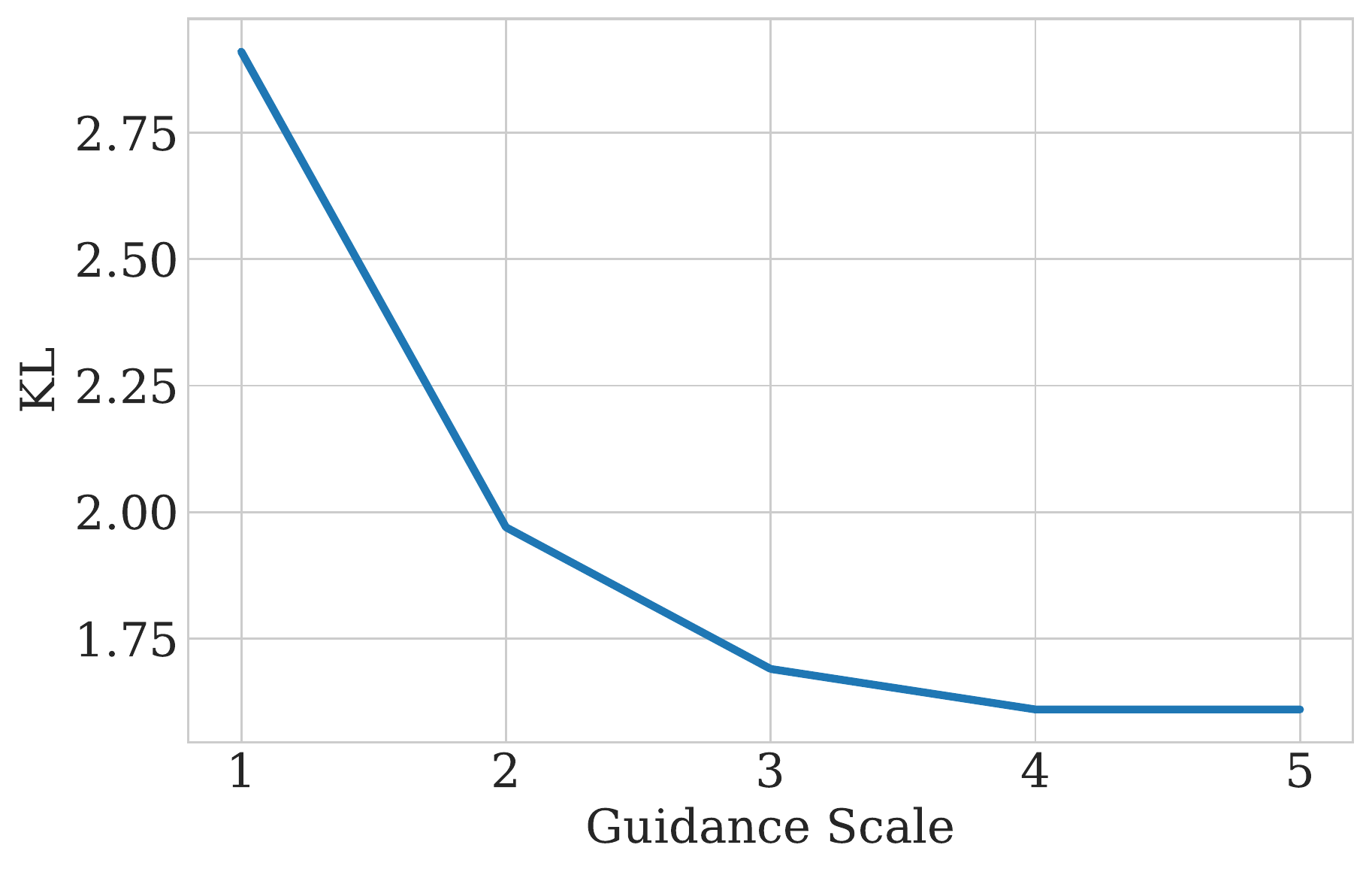}
    \label{fig:second}
\end{subfigure}
\caption{\label{fig:cfg} Results of FAD (left) and KL (right) as a function of the guidance scale.}
\label{fig:figures}
\end{figure}

{\noindent \bf{The effect of classifier-free guidance scale.}} As pointed out by~\citet{ho2021classifier}, the choice of the \ac{CFG} scale offers a trade-off between sample diversity and quality with respect to the conditioning text. To gauge the effect of the $\gamma$ parameter in our setting on Figure~\ref{fig:cfg} we report results for $\gamma \in \{1.0, 2.0, 3.0\}$. Notice, setting $\gamma=1$ is equivalent to a vanilla sampling procedure (no \ac{CFG}).

Removing the \ac{CFG} results in poor performance compared to $\gamma > 1.0$. The FAD score reaches its' minimum at $\gamma = 3.0$, while the KL monotonically decreases and converges at $\gamma=4.0$. This implies that using $\gamma = 3.0$ provides the best trade-off in the evaluated setting between quality and diversity.

\begin{table}[t!]
\centering \small
\caption{\label{tab:mst} Multi-stream results: we report two sets of results encoding and generation. For the encoding metrics we report Bitrate (kbps), SI-SNR (dBs), and ViSQOL. For the generation scores we report FAD, KL and inference speed-up. We additionally include the number of streams and the Down Sampling Factor (DSF). Notice, the encoding metrics are the same for the large and base model as the same audio representation model was used.
}
\resizebox{1.0\columnwidth}{!}{%
\begin{tabular}{l|cc|ccc|cccc}
\toprule
 & & & \multicolumn{3}{c}{\textsc{Encoding}} & \multicolumn{3}{c}{\textsc{Generation}} & \\
 & \textsc{\# streams} & \textsc{DSF} & Bitrate (kbps) & \textsc{SI-SNR}$\uparrow$ & \textsc{ViSQOL}$\uparrow$ & \textsc{FAD}$\downarrow$ & \textsc{KL}$\downarrow$ & \textsc{Speed-Up}\\
\midrule
\multirow{3}{*}{\rotatebox{90}{base}} & 1 & x32 & 5.37 & 5.1 & 3.95 & 3.13  & 2.09 & x1.0\\
& 2 & x64 & 4.88 & 4.5 & 3.94 & 10.35 & 2.17 & x2.0 \\
& 4 & x128 & 4.39 & 4.2 & 3.91 & 9.68  & 2.36 & x5.1 \\
\midrule
\multirow{3}{*}{\rotatebox{90}{large}} & 1 & x32 & 5.37 & 5.1 & 3.95 & 1.82 & 1.69 & x1.0\\
& 2 & x64 & 4.88 & 4.5 & 3.94 & 6.89  & 1.86 & x2.3 \\
& 4 & x128 & 4.39 & 4.2 & 3.91 & 10.89 & 2.59 & x3.6 \\
\bottomrule
\end{tabular}}
\end{table}

{\noindent \bf{Multi-stream processing.}} To better understand the benefit of using multiple streams over a single stream we optimized three different audio discrete representation models using Down Sampling Factors (DSF) of \{x32, x64, x128\} using \{1, 2, 4\} different codebooks respectively. For a fair comparison we kept all models at 2048 codes overall (e.g., 2 codebooks of size 1024 for the x64 model and 4 codebooks of size 512 for the x128 model). 

To assess the quality of the audio representation we report three encoding metrics namely \ac{SI-SNR}, ViSQOL~\citep{chinen2020visqol}, and bitrate. These metrics characterize the audio representation only while ignoring the \ac{ALM}. Note, both \ac{SI-SNR} and ViSQOL are reference based metrics hence are computed between the audio reconstructed from the learned representation and the original audio sequence. Results are reported on Table~\ref{tab:mst}. While a single stream encoder achieves better \ac{SI-SNR} results, all representations are comparable in terms of estimated perceptual quality as measured by ViSQOL. While the total amount of codes is the same across all settings, the effective bitrate of the multi-stream models was degraded, leading to lower \ac{SI-SNR} values. 

Next, we report the FAD, KL and inference speed-up for the \ac{ALM} on top of the learned representations. Results suggest that increasing the number of streams degrades the performance in both base and large models when compared to a single-stream. 
\audiogen-base improves the KL score over DiffSound (2.57) while yielding higher FAD. \audiogen-large with two streams improves over DiffSound in both FAD and KL (7.39 and 2.57 respectively). While showing inferior FAD and KL results when compared to the single-stream model, the multi-stream setting offers a trade-off between inference quality and speed.

\section{Limitations}
\label{sec:lim}
As we operate on audio tokens using relatively small down-sampling factor, the audio tokens sequences can be extremely long. This impose two main limitations: (i) modeling long range sequences; (ii) high inference time. In this work, we propose one possible relaxation to the first limitation, however such an approach comes at the cost of producing less quality audio samples. Such issues will become worse when considering high resolution audio samples, (e.g., sampling rate of 48kHz). Another limitation of the proposed approach relates to the audio compositions. Although, the mixing augmentation greatly improves the models ability to separate the sources and create complex compositions, it is still lacks the understanding of temporal ordering in the scene, e.g., a dog is barking \textbf{then} a birds humming, vs. a dog is barking and a birds humming in the \textbf{background}. Lastly, as we omit most of the speech samples in our training set, the proposed approach often generates unintelligible speech. This can be mitigated by either using more speech data, better data augmentation recipes for speech, or by providing additional speech features. Another limitation of the datasets used is its diversity. The datasets were mainly collected from YouTube, in which specific demographic and geographic locations are more represent than others. This may create bias in the generated sample. 

\section{Conclusion}
\label{sec:con}
In this work, we propose a Transformer based generative model, named \audiogen, which operates on a learnt discrete audio representation. Unlike previous work, we empirically demonstrate that auto-regressive models can generate high-quality audio samples conditionally or unconditionally. We show that on-the-fly text and audio mixing augmentations can improve model performance and provide an ablation study, analyzing the effect of \ac{CFG} and multi-stream processing. 

As for broader impacts, this work serves as the foundation for building better text-to-audio models. In addition, the proposed research could open up future directions involved with benchmarking, semantic audio editing, audio source separation from discrete units, etc.

\bibliography{iclr2023_conference}

\begin{thebibliography}{69}
\providecommand{\natexlab}[1]{#1}
\providecommand{\url}[1]{\texttt{#1}}
\expandafter\ifx\csname urlstyle\endcsname\relax
  \providecommand{\doi}[1]{doi: #1}\else
  \providecommand{\doi}{doi: \begingroup \urlstyle{rm}\Url}\fi

\bibitem[Ao et~al.(2022)Ao, Wang, Zhou, Wang, Ren, Wu, Liu, Ko, Li, Zhang,
  et~al.]{ao2021speecht5}
Junyi Ao, Rui Wang, Long Zhou, Chengyi Wang, Shuo Ren, Yu~Wu, Shujie Liu, Tom
  Ko, Qing Li, Yu~Zhang, et~al.
\newblock Speecht5: Unified-modal encoder-decoder pre-training for spoken
  language processing.
\newblock In \emph{Proceedings of the 60th Annual Meeting of the Association
  for Computational Linguistics (Volume 1: Long Papers)}, pp.\  5723--5738,
  2022.

\bibitem[Ba et~al.(2016)Ba, Kiros, and Hinton]{ba2016layer}
Jimmy~Lei Ba, Jamie~Ryan Kiros, and Geoffrey~E Hinton.
\newblock Layer normalization.
\newblock \emph{arXiv preprint arXiv:1607.06450}, 2016.

\bibitem[Baevski et~al.(2020)]{baevski2020wav2vec}
Alexei Baevski et~al.
\newblock wav2vec 2.0: A framework for self-supervised learning of speech
  representations.
\newblock In \emph{ICLR}, 2020.

\bibitem[Beltagy et~al.(2020)Beltagy, Peters, and Cohan]{beltagy2020longformer}
Iz~Beltagy, Matthew~E Peters, and Arman Cohan.
\newblock Longformer: The long-document transformer.
\newblock \emph{arXiv preprint arXiv:2004.05150}, 2020.

\bibitem[Bird et~al.(2009)Bird, Klein, and Loper]{bird2009natural}
Steven Bird, Ewan Klein, and Edward Loper.
\newblock \emph{Natural language processing with Python: analyzing text with
  the natural language toolkit}.
\newblock " O'Reilly Media, Inc.", 2009.

\bibitem[Borsos et~al.(2022)Borsos, Marinier, Vincent, Kharitonov, Pietquin,
  Sharifi, Teboul, Grangier, Tagliasacchi, and Zeghidour]{borsos2022audiolm}
Zal{\'a}n Borsos, Rapha{\"e}l Marinier, Damien Vincent, Eugene Kharitonov,
  Olivier Pietquin, Matt Sharifi, Olivier Teboul, David Grangier, Marco
  Tagliasacchi, and Neil Zeghidour.
\newblock Audiolm: a language modeling approach to audio generation.
\newblock \emph{arXiv preprint arXiv:2209.03143}, 2022.

\bibitem[Brown et~al.(2020)Brown, Mann, Ryder, Subbiah, Kaplan, Dhariwal,
  Neelakantan, Shyam, Sastry, Askell, et~al.]{brown2020language}
Tom Brown, Benjamin Mann, Nick Ryder, Melanie Subbiah, Jared~D Kaplan, Prafulla
  Dhariwal, Arvind Neelakantan, Pranav Shyam, Girish Sastry, Amanda Askell,
  et~al.
\newblock Language models are few-shot learners.
\newblock \emph{Advances in neural information processing systems},
  33:\penalty0 1877--1901, 2020.

\bibitem[Capon(1969)]{capon1969high}
Jack Capon.
\newblock High-resolution frequency-wavenumber spectrum analysis.
\newblock \emph{Proceedings of the IEEE}, 57\penalty0 (8):\penalty0 1408--1418,
  1969.

\bibitem[Chen et~al.(2020)Chen, Xie, Vedaldi, and Zisserman]{chen2020vggsound}
Honglie Chen, Weidi Xie, Andrea Vedaldi, and Andrew Zisserman.
\newblock Vggsound: A large-scale audio-visual dataset.
\newblock In \emph{ICASSP 2020-2020 IEEE International Conference on Acoustics,
  Speech and Signal Processing (ICASSP)}, pp.\  721--725. IEEE, 2020.

\bibitem[Chen et~al.(2022)Chen, Wang, Chen, Wu, Liu, Chen, Li, Kanda, Yoshioka,
  Xiao, et~al.]{chen2022wavlm}
Sanyuan Chen, Chengyi Wang, Zhengyang Chen, Yu~Wu, Shujie Liu, Zhuo Chen, Jinyu
  Li, Naoyuki Kanda, Takuya Yoshioka, Xiong Xiao, et~al.
\newblock Wavlm: Large-scale self-supervised pre-training for full stack speech
  processing.
\newblock \emph{IEEE Journal of Selected Topics in Signal Processing}, 2022.

\bibitem[Chinen et~al.(2020)Chinen, Lim, Skoglund, Gureev, O'Gorman, and
  Hines]{chinen2020visqol}
Michael Chinen, Felicia~SC Lim, Jan Skoglund, Nikita Gureev, Feargus O'Gorman,
  and Andrew Hines.
\newblock Visqol v3: An open source production ready objective speech and audio
  metric.
\newblock In \emph{2020 twelfth international conference on quality of
  multimedia experience (QoMEX)}, pp.\  1--6. IEEE, 2020.

\bibitem[Clevert et~al.(2015)Clevert, Unterthiner, and
  Hochreiter]{clevert2015fast}
Djork-Arn{\'e} Clevert, Thomas Unterthiner, and Sepp Hochreiter.
\newblock Fast and accurate deep network learning by exponential linear units
  (elus).
\newblock \emph{arXiv preprint arXiv:1511.07289}, 2015.

\bibitem[Devlin et~al.(2019)]{devlin-etal-2019-bert}
Jacob Devlin et~al.
\newblock {BERT}: Pre-training of deep bidirectional transformers for language
  understanding.
\newblock In \emph{NAACL}, 2019.

\bibitem[Drossos et~al.(2020)Drossos, Lipping, and Virtanen]{drossos2020clotho}
Konstantinos Drossos, Samuel Lipping, and Tuomas Virtanen.
\newblock Clotho: An audio captioning dataset.
\newblock In \emph{ICASSP 2020-2020 IEEE International Conference on Acoustics,
  Speech and Signal Processing (ICASSP)}, pp.\  736--740. IEEE, 2020.

\bibitem[Ebbers et~al.(2017)]{ebbers2017hidden}
Janek Ebbers et~al.
\newblock Hidden markov model variational autoencoder for acoustic unit
  discovery.
\newblock In \emph{INTERSPEECH 2017}, 2017.

\bibitem[Fonseca et~al.(2021)Fonseca, Favory, Pons, Font, and
  Serra]{fonseca2021fsd50k}
Eduardo Fonseca, Xavier Favory, Jordi Pons, Frederic Font, and Xavier Serra.
\newblock Fsd50k: an open dataset of human-labeled sound events.
\newblock \emph{IEEE/ACM Transactions on Audio, Speech, and Language
  Processing}, 30:\penalty0 829--852, 2021.

\bibitem[Frost(1972)]{frost1972algorithm}
Otis~Lamont Frost.
\newblock An algorithm for linearly constrained adaptive array processing.
\newblock \emph{Proceedings of the IEEE}, 60\penalty0 (8):\penalty0 926--935,
  1972.

\bibitem[Gafni et~al.(2022)Gafni, Polyak, Ashual, Sheynin, Parikh, and
  Taigman]{gafni2022make}
Oran Gafni, Adam Polyak, Oron Ashual, Shelly Sheynin, Devi Parikh, and Yaniv
  Taigman.
\newblock Make-a-scene: Scene-based text-to-image generation with human priors.
\newblock In \emph{Computer Vision--ECCV 2022: 17th European Conference, Tel
  Aviv, Israel, October 23--27, 2022, Proceedings, Part XV}, pp.\  89--106.
  Springer, 2022.

\bibitem[Gemmeke et~al.(2017)Gemmeke, Ellis, Freedman, Jansen, Lawrence, Moore,
  Plakal, and Ritter]{gemmeke2017audio}
Jort~F Gemmeke, Daniel~PW Ellis, Dylan Freedman, Aren Jansen, Wade Lawrence,
  R~Channing Moore, Manoj Plakal, and Marvin Ritter.
\newblock Audio set: An ontology and human-labeled dataset for audio events.
\newblock In \emph{2017 IEEE international conference on acoustics, speech and
  signal processing (ICASSP)}, pp.\  776--780. IEEE, 2017.

\bibitem[Gritsenko et~al.(2020)Gritsenko, Salimans, van~den Berg, Snoek, and
  Kalchbrenner]{gritsenko2020spectral}
Alexey Gritsenko, Tim Salimans, Rianne van~den Berg, Jasper Snoek, and Nal
  Kalchbrenner.
\newblock A spectral energy distance for parallel speech synthesis.
\newblock \emph{Advances in Neural Information Processing Systems},
  33:\penalty0 13062--13072, 2020.

\bibitem[Ho \& Salimans(2021)Ho and Salimans]{ho2021classifier}
Jonathan Ho and Tim Salimans.
\newblock Classifier-free diffusion guidance.
\newblock In \emph{NeurIPS 2021 Workshop on Deep Generative Models and
  Downstream Applications}, 2021.

\bibitem[Holtzman et~al.(2019)Holtzman, Buys, Du, Forbes, and
  Choi]{holtzman2019curious}
Ari Holtzman, Jan Buys, Li~Du, Maxwell Forbes, and Yejin Choi.
\newblock The curious case of neural text degeneration.
\newblock In \emph{International Conference on Learning Representations}, 2019.

\bibitem[Hsu et~al.(2017)Hsu, Zhang, and Glass]{hsu2017unsupervised}
Wei-Ning Hsu, Yu~Zhang, and James Glass.
\newblock Unsupervised learning of disentangled and interpretable
  representations from sequential data.
\newblock In \emph{Advances in Neural Information Processing Systems}, 2017.

\bibitem[Hsu et~al.(2021)Hsu, Bolte, Tsai, Lakhotia, Salakhutdinov, and
  Mohamed]{hsu2021hubert}
Wei-Ning Hsu, Benjamin Bolte, Yao-Hung~Hubert Tsai, Kushal Lakhotia, Ruslan
  Salakhutdinov, and Abdelrahman Mohamed.
\newblock Hubert: Self-supervised speech representation learning by masked
  prediction of hidden units.
\newblock \emph{IEEE/ACM Transactions on Audio, Speech, and Language
  Processing}, 29:\penalty0 3451--3460, 2021.

\bibitem[Huang et~al.(2022)Huang, Xu, Li, Baevski, Auli, Galuba, Metze, and
  Feichtenhofer]{xu2022masked}
Po-Yao Huang, Hu~Xu, Juncheng~B Li, Alexei Baevski, Michael Auli, Wojciech
  Galuba, Florian Metze, and Christoph Feichtenhofer.
\newblock Masked autoencoders that listen.
\newblock In Alice~H. Oh, Alekh Agarwal, Danielle Belgrave, and Kyunghyun Cho
  (eds.), \emph{Advances in Neural Information Processing Systems}, 2022.
\newblock URL \url{https://openreview.net/forum?id=MAMOi89bOL}.

\bibitem[Karras et~al.(2019)Karras, Laine, and Aila]{karras2019style}
Tero Karras, Samuli Laine, and Timo Aila.
\newblock A style-based generator architecture for generative adversarial
  networks.
\newblock In \emph{Proceedings of the IEEE/CVF conference on computer vision
  and pattern recognition}, pp.\  4401--4410, 2019.

\bibitem[Kharitonov et~al.(2022{\natexlab{a}})Kharitonov, Copet, Lakhotia,
  Nguyen, Tomasello, Lee, Elkahky, Hsu, Mohamed, Dupoux,
  et~al.]{kharitonov2022textless}
Eugene Kharitonov, Jade Copet, Kushal Lakhotia, Tu~Anh Nguyen, Paden Tomasello,
  Ann Lee, Ali Elkahky, Wei-Ning Hsu, Abdelrahman Mohamed, Emmanuel Dupoux,
  et~al.
\newblock textless-lib: a library for textless spoken language processing.
\newblock In \emph{Proceedings of the 2022 Conference of the North American
  Chapter of the Association for Computational Linguistics: Human Language
  Technologies: System Demonstrations}, pp.\  1--9, 2022{\natexlab{a}}.

\bibitem[Kharitonov et~al.(2022{\natexlab{b}})Kharitonov, Lee, Polyak, Adi,
  Copet, Lakhotia, Nguyen, Riviere, Mohamed, Dupoux,
  et~al.]{kharitonov2021text}
Eugene Kharitonov, Ann Lee, Adam Polyak, Yossi Adi, Jade Copet, Kushal
  Lakhotia, Tu~Anh Nguyen, Morgane Riviere, Abdelrahman Mohamed, Emmanuel
  Dupoux, et~al.
\newblock Text-free prosody-aware generative spoken language modeling.
\newblock In \emph{Proceedings of the 60th Annual Meeting of the Association
  for Computational Linguistics (Volume 1: Long Papers)}, pp.\  8666--8681,
  2022{\natexlab{b}}.

\bibitem[Kilgour et~al.(2019)Kilgour, Zuluaga, Roblek, and
  Sharifi]{Kilgour2019FrchetAD}
Kevin Kilgour, Mauricio Zuluaga, Dominik Roblek, and Matthew Sharifi.
\newblock Fr{\'e}chet audio distance: A reference-free metric for evaluating
  music enhancement algorithms.
\newblock In \emph{INTERSPEECH}, 2019.

\bibitem[Kim et~al.(2019)Kim, Kim, Lee, and Kim]{kim2019audiocaps}
Chris~Dongjoo Kim, Byeongchang Kim, Hyunmin Lee, and Gunhee Kim.
\newblock Audiocaps: Generating captions for audios in the wild.
\newblock In \emph{Proceedings of the 2019 Conference of the North American
  Chapter of the Association for Computational Linguistics: Human Language
  Technologies, Volume 1 (Long and Short Papers)}, pp.\  119--132, 2019.

\bibitem[Kong et~al.(2020)Kong, Kim, and Bae]{kong2020hifi}
Jungil Kong, Jaehyeon Kim, and Jaekyoung Bae.
\newblock Hifi-gan: Generative adversarial networks for efficient and high
  fidelity speech synthesis.
\newblock \emph{Advances in Neural Information Processing Systems},
  33:\penalty0 17022--17033, 2020.

\bibitem[Koutini et~al.(2021)Koutini, Schl{\"u}ter, Eghbal-zadeh, and
  Widmer]{koutini2021efficient}
Khaled Koutini, Jan Schl{\"u}ter, Hamid Eghbal-zadeh, and Gerhard Widmer.
\newblock Efficient training of audio transformers with patchout.
\newblock \emph{arXiv preprint arXiv:2110.05069}, 2021.

\bibitem[Kreuk et~al.(2020)Kreuk, Keshet, and Adi]{kreuk2020self}
Felix Kreuk, Joseph Keshet, and Yossi Adi.
\newblock Self-supervised contrastive learning for unsupervised phoneme
  segmentation.
\newblock In \emph{INTERSPEECH 2020}, 2020.

\bibitem[Kreuk et~al.(2022)Kreuk, Polyak, Copet, Kharitonov, Nguyen,
  Rivi{\`e}re, Hsu, Mohamed, Dupoux, and Adi]{kreuk2021textless}
Felix Kreuk, Adam Polyak, Jade Copet, Eugene Kharitonov, Tu~Anh Nguyen, Morgan
  Rivi{\`e}re, Wei-Ning Hsu, Abdelrahman Mohamed, Emmanuel Dupoux, and Yossi
  Adi.
\newblock Textless speech emotion conversion using discrete \& decomposed
  representations.
\newblock In \emph{Proceedings of the 2022 Conference on Empirical Methods in
  Natural Language Processing}, pp.\  11200--11214, 2022.

\bibitem[Kumar et~al.(2019)Kumar, Kumar, de~Boissiere, Gestin, Teoh, Sotelo,
  de~Br{\'e}bisson, Bengio, and Courville]{melgan}
Kundan Kumar, Rithesh Kumar, Thibault de~Boissiere, Lucas Gestin, Wei~Zhen
  Teoh, Jose Sotelo, Alexandre de~Br{\'e}bisson, Yoshua Bengio, and Aaron~C
  Courville.
\newblock Melgan: Generative adversarial networks for conditional waveform
  synthesis.
\newblock \emph{Advances in neural information processing systems}, 32, 2019.

\bibitem[Lakhotia et~al.(2021)Lakhotia, Kharitonov, Hsu, Adi, Polyak, Bolte,
  Nguyen, Copet, Baevski, Mohamed, et~al.]{lakhotia2021generative}
Kushal Lakhotia, Eugene Kharitonov, Wei-Ning Hsu, Yossi Adi, Adam Polyak,
  Benjamin Bolte, Tu-Anh Nguyen, Jade Copet, Alexei Baevski, Abdelrahman
  Mohamed, et~al.
\newblock On generative spoken language modeling from raw audio.
\newblock \emph{Transactions of the Association for Computational Linguistics},
  9:\penalty0 1336--1354, 2021.

\bibitem[Lee et~al.(2022{\natexlab{a}})Lee, Chen, Wang, Gu, Popuri, Ma, Polyak,
  Adi, He, Tang, et~al.]{lee2021direct}
Ann Lee, Peng-Jen Chen, Changhan Wang, Jiatao Gu, Sravya Popuri, Xutai Ma, Adam
  Polyak, Yossi Adi, Qing He, Yun Tang, et~al.
\newblock Direct speech-to-speech translation with discrete units.
\newblock In \emph{Proceedings of the 60th Annual Meeting of the Association
  for Computational Linguistics (Volume 1: Long Papers)}, pp.\  3327--3339,
  2022{\natexlab{a}}.

\bibitem[Lee et~al.(2022{\natexlab{b}})Lee, Gong, Duquenne, Schwenk, Chen,
  Wang, Popuri, Adi, Pino, Gu, and Hsu]{lee-etal-2022-textless}
Ann Lee, Hongyu Gong, Paul-Ambroise Duquenne, Holger Schwenk, Peng-Jen Chen,
  Changhan Wang, Sravya Popuri, Yossi Adi, Juan Pino, Jiatao Gu, and Wei-Ning
  Hsu.
\newblock Textless speech-to-speech translation on real data.
\newblock In \emph{Proceedings of the 2022 Conference of the North American
  Chapter of the Association for Computational Linguistics: Human Language
  Technologies}, pp.\  860--872, Seattle, United States, July
  2022{\natexlab{b}}. Association for Computational Linguistics.
\newblock \doi{10.18653/v1/2022.naacl-main.63}.
\newblock URL \url{https://aclanthology.org/2022.naacl-main.63}.

\bibitem[Li et~al.(2021)Li, Tagliasacchi, Rybakov, Ungureanu, and
  Roblek]{li2021real}
Yunpeng Li, Marco Tagliasacchi, Oleg Rybakov, Victor Ungureanu, and Dominik
  Roblek.
\newblock Real-time speech frequency bandwidth extension.
\newblock In \emph{ICASSP 2021-2021 IEEE International Conference on Acoustics,
  Speech and Signal Processing (ICASSP)}, pp.\  691--695. IEEE, 2021.

\bibitem[Nguyen et~al.(2022)Nguyen, Kharitonov, Copet, Adi, Hsu, Elkahky,
  Tomasello, Algayres, Sagot, Mohamed, et~al.]{nguyen2022generative}
Tu~Anh Nguyen, Eugene Kharitonov, Jade Copet, Yossi Adi, Wei-Ning Hsu, Ali
  Elkahky, Paden Tomasello, Robin Algayres, Benoit Sagot, Abdelrahman Mohamed,
  et~al.
\newblock Generative spoken dialogue language modeling.
\newblock \emph{arXiv preprint arXiv:2203.16502}, 2022.

\bibitem[Nichol et~al.(2022)Nichol, Dhariwal, Ramesh, Shyam, Mishkin, Mcgrew,
  Sutskever, and Chen]{nichol2021glide}
Alexander~Quinn Nichol, Prafulla Dhariwal, Aditya Ramesh, Pranav Shyam, Pamela
  Mishkin, Bob Mcgrew, Ilya Sutskever, and Mark Chen.
\newblock Glide: Towards photorealistic image generation and editing with
  text-guided diffusion models.
\newblock In \emph{International Conference on Machine Learning}, pp.\
  16784--16804. PMLR, 2022.

\bibitem[Oord et~al.(2016)Oord, Dieleman, Zen, Simonyan, Vinyals, Graves,
  Kalchbrenner, Senior, and Kavukcuoglu]{oord2016wavenet}
Aaron van~den Oord, Sander Dieleman, Heiga Zen, Karen Simonyan, Oriol Vinyals,
  Alex Graves, Nal Kalchbrenner, Andrew Senior, and Koray Kavukcuoglu.
\newblock Wavenet: A generative model for raw audio.
\newblock \emph{arXiv preprint arXiv:1609.03499}, 2016.

\bibitem[Park et~al.(2019)Park, Liu, Wang, and Zhu]{park2019semantic}
Taesung Park, Ming-Yu Liu, Ting-Chun Wang, and Jun-Yan Zhu.
\newblock Semantic image synthesis with spatially-adaptive normalization.
\newblock In \emph{Proceedings of the IEEE/CVF conference on computer vision
  and pattern recognition}, pp.\  2337--2346, 2019.

\bibitem[Polyak et~al.(2021)Polyak, Adi, Copet, Kharitonov, Lakhotia, Hsu,
  Mohamed, and Dupoux]{polyak2021speech}
Adam Polyak, Yossi Adi, Jade Copet, Eugene Kharitonov, Kushal Lakhotia,
  Wei-Ning Hsu, Abdelrahman Mohamed, and Emmanuel Dupoux.
\newblock Speech resynthesis from discrete disentangled self-supervised
  representations.
\newblock In \emph{INTERSPEECH}, 2021.

\bibitem[Popuri et~al.(2022)Popuri, Chen, Wang, Pino, Adi, Gu, Hsu, and
  Lee]{popuri2022enhanced}
Sravya Popuri, Peng-Jen Chen, Changhan Wang, Juan Pino, Yossi Adi, Jiatao Gu,
  Wei-Ning Hsu, and Ann Lee.
\newblock {Enhanced Direct Speech-to-Speech Translation Using Self-supervised
  Pre-training and Data Augmentation}.
\newblock In \emph{Proc. Interspeech 2022}, pp.\  5195--5199, 2022.
\newblock \doi{10.21437/Interspeech.2022-11032}.

\bibitem[Radford et~al.(2019)Radford, Wu, Child, Luan, Amodei, and
  Sutskever]{radford2019language}
Alec Radford, Jeff Wu, Rewon Child, David Luan, Dario Amodei, and Ilya
  Sutskever.
\newblock Language models are unsupervised multitask learners.
\newblock 2019.

\bibitem[Rae et~al.(2020)Rae, Potapenko, Jayakumar, Hillier, and
  Lillicrap]{rae2019compressive}
Jack~W Rae, Anna Potapenko, Siddhant~M Jayakumar, Chloe Hillier, and Timothy~P
  Lillicrap.
\newblock Compressive transformers for long-range sequence modelling.
\newblock In \emph{International Conference on Learning Representations}, 2020.

\bibitem[Raffel et~al.(2020)Raffel, Shazeer, Roberts, Lee, Narang, Matena,
  Zhou, Li, Liu, et~al.]{raffel2020exploring}
Colin Raffel, Noam Shazeer, Adam Roberts, Katherine Lee, Sharan Narang, Michael
  Matena, Yanqi Zhou, Wei Li, Peter~J Liu, et~al.
\newblock Exploring the limits of transfer learning with a unified text-to-text
  transformer.
\newblock \emph{J. Mach. Learn. Res.}, 21\penalty0 (140):\penalty0 1--67, 2020.

\bibitem[Ramesh et~al.(2022)Ramesh, Dhariwal, Nichol, Chu, and
  Chen]{ramesh2022hierarchical}
Aditya Ramesh, Prafulla Dhariwal, Alex Nichol, Casey Chu, and Mark Chen.
\newblock Hierarchical text-conditional image generation with clip latents.
\newblock \emph{arXiv preprint arXiv:2204.06125}, 2022.

\bibitem[Reddy et~al.(2021)Reddy, Basha, Hari, and Penchalaiah]{dalle}
Mr~D~Murahari Reddy, Mr~Sk~Masthan Basha, Mr~M~Chinnaiahgari Hari, and Mr~N
  Penchalaiah.
\newblock Dall-e: Creating images from text.
\newblock \emph{UGC Care Group I Journal}, 8\penalty0 (14):\penalty0 71--75,
  2021.

\bibitem[Ren et~al.(2021)Ren, Hu, Tan, Qin, Zhao, Zhao, and
  Liu]{ren2020fastspeech}
Yi~Ren, Chenxu Hu, Xu~Tan, Tao Qin, Sheng Zhao, Zhou Zhao, and Tie-Yan Liu.
\newblock Fastspeech 2: Fast and high-quality end-to-end text to speech.
\newblock In \emph{International Conference on Learning Representations}, 2021.

\bibitem[{Ribeiro} et~al.(2011){Ribeiro}, {Flor{\^e}ncio}, {Zhang}, and
  {Seltzer}]{ribeiro2011crowdmos}
F.~{Ribeiro}, D.~{Flor{\^e}ncio}, C.~{Zhang}, and M.~{Seltzer}.
\newblock {CROWDMOS}: An approach for crowdsourcing mean opinion score studies.
\newblock In \emph{IEEE International Conference on Acoustics, Speech and
  Signal Processing (ICASSP)}, pp.\  2416--2419, 2011.
\newblock \doi{10.1109/ICASSP.2011.5946971}.

\bibitem[Rombach et~al.(2022)Rombach, Blattmann, Lorenz, Esser, and
  Ommer]{Rombach_2022_CVPR}
Robin Rombach, Andreas Blattmann, Dominik Lorenz, Patrick Esser, and Bj\"orn
  Ommer.
\newblock High-resolution image synthesis with latent diffusion models.
\newblock In \emph{Proceedings of the IEEE/CVF Conference on Computer Vision
  and Pattern Recognition (CVPR)}, pp.\  10684--10695, June 2022.

\bibitem[Saharia et~al.()Saharia, Chan, Saxena, Li, Whang, Denton, Ghasemipour,
  Gontijo-Lopes, Ayan, Salimans, et~al.]{imagen}
Chitwan Saharia, William Chan, Saurabh Saxena, Lala Li, Jay Whang, Emily
  Denton, Seyed Kamyar~Seyed Ghasemipour, Raphael Gontijo-Lopes, Burcu~Karagol
  Ayan, Tim Salimans, et~al.
\newblock Photorealistic text-to-image diffusion models with deep language
  understanding.
\newblock In \emph{Advances in Neural Information Processing Systems}.

\bibitem[Schneider et~al.(2019)Schneider, Baevski, Collobert, and
  Auli]{schneider2019wav2vec}
Steffen Schneider, Alexei Baevski, Ronan Collobert, and Michael Auli.
\newblock {wav2vec: Unsupervised Pre-Training for Speech Recognition}.
\newblock In \emph{INTERSPEECH}, 2019.

\bibitem[Series(2014)]{series2014method}
B~Series.
\newblock Method for the subjective assessment of intermediate quality level of
  audio systems.
\newblock \emph{International Telecommunication Union Radiocommunication
  Assembly}, 2014.

\bibitem[Suzuki et~al.(2003)Suzuki, Mellert, Richter, M{\o}ller, Nielsen,
  Hellman, Ashihara, Ozawa, and Takeshima]{suzuki2003precise}
Y{\^o}iti Suzuki, Volker Mellert, Utz Richter, Henrik M{\o}ller, Leif Nielsen,
  Rhona Hellman, Kaoru Ashihara, Kenji Ozawa, and Hisashi Takeshima.
\newblock Precise and full-range determination of two-dimensional equal
  loudness contours.
\newblock \emph{Tohoku University, Japan}, 2003.

\bibitem[van~den Oord et~al.(2017{\natexlab{a}})van~den Oord, Vinyals, and
  kavukcuoglu]{NIPS2017_7a98af17}
Aaron van~den Oord, Oriol Vinyals, and koray kavukcuoglu.
\newblock Neural discrete representation learning.
\newblock In I.~Guyon, U.~Von Luxburg, S.~Bengio, H.~Wallach, R.~Fergus,
  S.~Vishwanathan, and R.~Garnett (eds.), \emph{Advances in Neural Information
  Processing Systems}, volume~30. Curran Associates, Inc., 2017{\natexlab{a}}.
\newblock URL
  \url{https://proceedings.neurips.cc/paper/2017/file/7a98af17e63a0ac09ce2e96d03992fbc-Paper.pdf}.

\bibitem[van~den Oord et~al.(2017{\natexlab{b}})van~den Oord, Vinyals,
  et~al.]{van2017neural}
Aaron van~den Oord, Oriol Vinyals, et~al.
\newblock Neural discrete representation learning.
\newblock In \emph{NeurIPS}, 2017{\natexlab{b}}.

\bibitem[van~den Oord et~al.(2018)van~den Oord, Li, and
  Vinyals]{oord2018representation}
Aaron van~den Oord, Yazhe Li, and Oriol Vinyals.
\newblock Representation learning with contrastive predictive coding.
\newblock \emph{arXiv preprint arXiv:1807.03748}, 2018.

\bibitem[Vasuki \& Vanathi(2006)Vasuki and Vanathi]{vasuki2006review}
A~Vasuki and PT~Vanathi.
\newblock A review of vector quantization techniques.
\newblock \emph{IEEE Potentials}, 25\penalty0 (4):\penalty0 39--47, 2006.

\bibitem[Wang et~al.(2021)Wang, Wu, Qian, Kumatani, Liu, Wei, Zeng, and
  Huang]{wang2021unispeech}
Chengyi Wang, Yu~Wu, Yao Qian, Kenichi Kumatani, Shujie Liu, Furu Wei, Michael
  Zeng, and Xuedong Huang.
\newblock Unispeech: Unified speech representation learning with labeled and
  unlabeled data.
\newblock In \emph{International Conference on Machine Learning}, pp.\
  10937--10947. PMLR, 2021.

\bibitem[Yamamoto et~al.(2020)Yamamoto, Song, and Kim]{yamamoto2020parallel}
Ryuichi Yamamoto, Eunwoo Song, and Jae-Min Kim.
\newblock Parallel wavegan: A fast waveform generation model based on
  generative adversarial networks with multi-resolution spectrogram.
\newblock In \emph{ICASSP 2020-2020 IEEE International Conference on Acoustics,
  Speech and Signal Processing (ICASSP)}, pp.\  6199--6203. IEEE, 2020.

\bibitem[Yang et~al.(2022)Yang, Yu, Wang, Wang, Weng, Zou, and
  Yu]{yang2022diffsound}
Dongchao Yang, Jianwei Yu, Helin Wang, Wen Wang, Chao Weng, Yuexian Zou, and
  Dong Yu.
\newblock Diffsound: Discrete diffusion model for text-to-sound generation.
\newblock \emph{arXiv preprint arXiv:2207.09983}, 2022.

\bibitem[You et~al.(2021)You, Kim, Nam, Hwang, and Chae]{you2021gan}
Jaeseong You, Dalhyun Kim, Gyuhyeon Nam, Geumbyeol Hwang, and Gyeongsu Chae.
\newblock Gan vocoder: Multi-resolution discriminator is all you need.
\newblock \emph{arXiv preprint arXiv:2103.05236}, 2021.

\bibitem[Yu et~al.()Yu, Xu, Koh, Luong, Baid, Wang, Vasudevan, Ku, Yang, Ayan,
  et~al.]{parti}
Jiahui Yu, Yuanzhong Xu, Jing~Yu Koh, Thang Luong, Gunjan Baid, Zirui Wang,
  Vijay Vasudevan, Alexander Ku, Yinfei Yang, Burcu~Karagol Ayan, et~al.
\newblock Scaling autoregressive models for content-rich text-to-image
  generation.
\newblock \emph{Transactions on Machine Learning Research}.

\bibitem[Zaheer et~al.(2020)Zaheer, Guruganesh, Dubey, Ainslie, Alberti,
  Ontanon, Pham, Ravula, Wang, Yang, et~al.]{zaheer2020big}
Manzil Zaheer, Guru Guruganesh, Kumar~Avinava Dubey, Joshua Ainslie, Chris
  Alberti, Santiago Ontanon, Philip Pham, Anirudh Ravula, Qifan Wang, Li~Yang,
  et~al.
\newblock Big bird: Transformers for longer sequences.
\newblock \emph{Advances in Neural Information Processing Systems},
  33:\penalty0 17283--17297, 2020.

\bibitem[Zeghidour et~al.(2021)Zeghidour, Luebs, Omran, Skoglund, and
  Tagliasacchi]{zeghidour2021soundstream}
Neil Zeghidour, Alejandro Luebs, Ahmed Omran, Jan Skoglund, and Marco
  Tagliasacchi.
\newblock Soundstream: An end-to-end neural audio codec.
\newblock \emph{IEEE/ACM Transactions on Audio, Speech, and Language
  Processing}, 2021.

\bibitem[Zhang et~al.(2022)Zhang, Roller, Goyal, Artetxe, Chen, Chen, Dewan,
  Diab, Li, Lin, et~al.]{zhang2022opt}
Susan Zhang, Stephen Roller, Naman Goyal, Mikel Artetxe, Moya Chen, Shuohui
  Chen, Christopher Dewan, Mona Diab, Xian Li, Xi~Victoria Lin, et~al.
\newblock Opt: Open pre-trained transformer language models.
\newblock \emph{arXiv preprint arXiv:2205.01068}, 2022.

\end{thebibliography}
\bibliographystyle{iclr2023_conference}

\clearpage
\section{Appendix}
\label{sec:app}

\subsection{Datasets}
We use a set of several datasets: AudioSet~\citep{gemmeke2017audio}, BBC sound effects~\footnote{\url{https://sound-effects.bbcrewind.co.uk/}}, AudioCaps~\citep{kim2019audiocaps}, Clotho v2~\citep{drossos2020clotho}, VGG-Sound~\citep{chen2020vggsound}, FSD50K~\citep{fonseca2021fsd50k}, Free To Use Sounds~\footnote{\url{https://www.freetousesounds.com/all-in-one-bundle/}}, Sonniss Game Effects~\footnote{\url{https://sonniss.com/gameaudiogdc}}, WeSoundEffects~\footnote{\url{https://wesoundeffects.com/we-sound-effects-bundle-2020/}}, Paramount Motion - Odeon Cinematic Sound Effects~\footnote{\url{https://www.paramountmotion.com/odeon-sound-effects}}. All audio files were sampled at 16kHz. 

\begin{table}[h!]
  \caption{Datasets description. Duration is reported in hours for original audio, i.e., before pre-processing. \label{tab:datasets}}
  \centering
  \resizebox{0.55\textwidth}{!}{
  \begin{tabular}{llrrrl}
    \toprule
    \textsc{Dataset} & \textsc{Text conditioning} & \textsc{Duration (h)}\\
    \midrule
    AudioSet & tags & 5.42k\\
    BBC & captions & 463\\
    AudioCaps & captions & 145\\
    Clotho v2 & captions & 37\\
    VGG-Sound & tags & 560\\ 
    FSD50K & tags \& captions & 108\\ 
    Free To Use Sounds & tags \& captions & 176\\ 
    Sonniss Game Effects & tags & 85\\ 
    WeSoundEffects & tags & 12\\ 
    Paramount Motion & tags & 20\\ 
    \bottomrule
  \end{tabular}}
\end{table}

\subsection{Additional Results}
\label{sec:app_res}

\begin{figure}[h!]
\centering
\begin{subfigure}{0.46\textwidth}
    \centering
    \includegraphics[width=\textwidth]{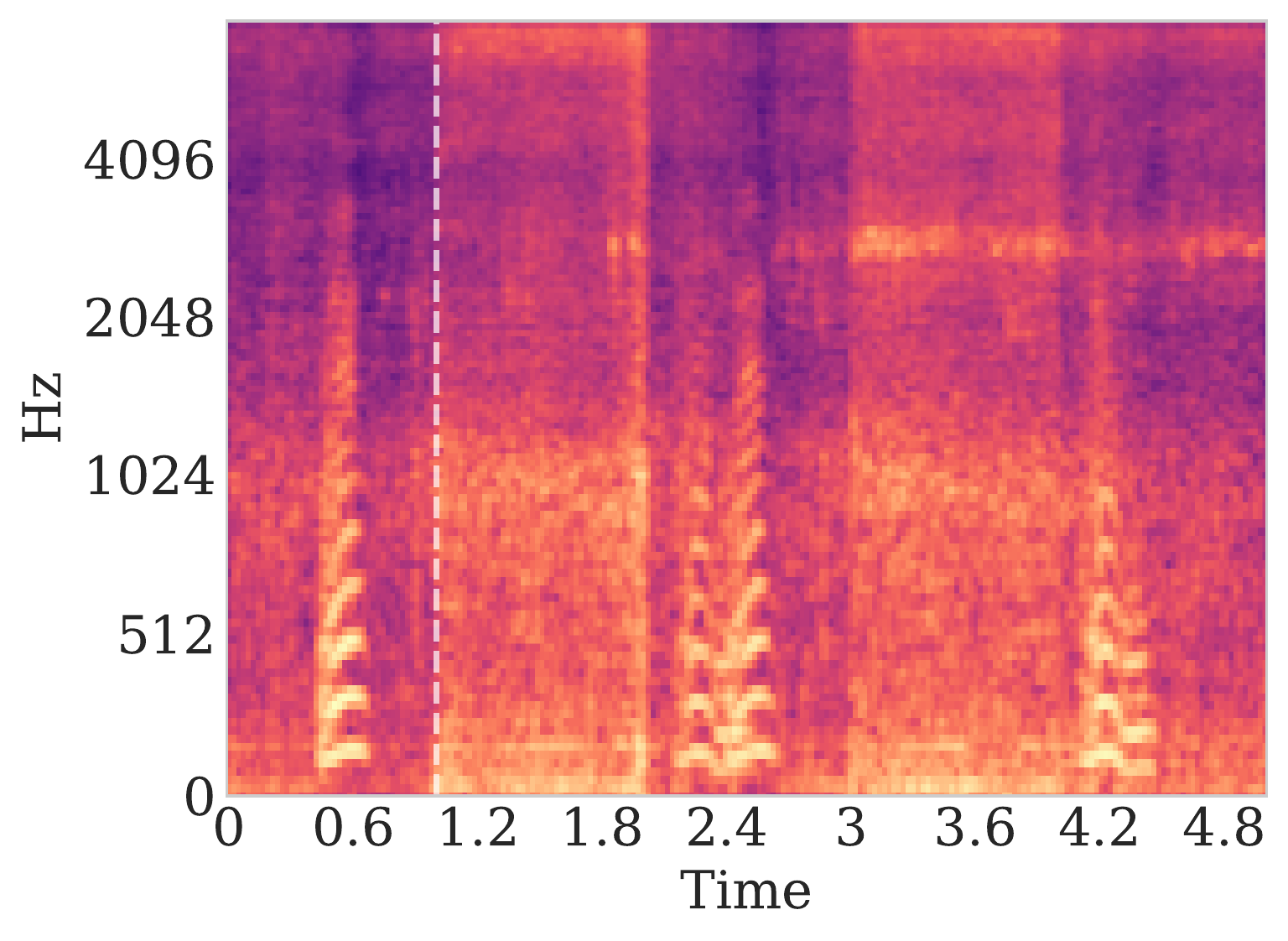}
    \label{fig:second}
\end{subfigure}
\hfill
\begin{subfigure}{0.46\textwidth}
    \centering
    \includegraphics[width=\textwidth]{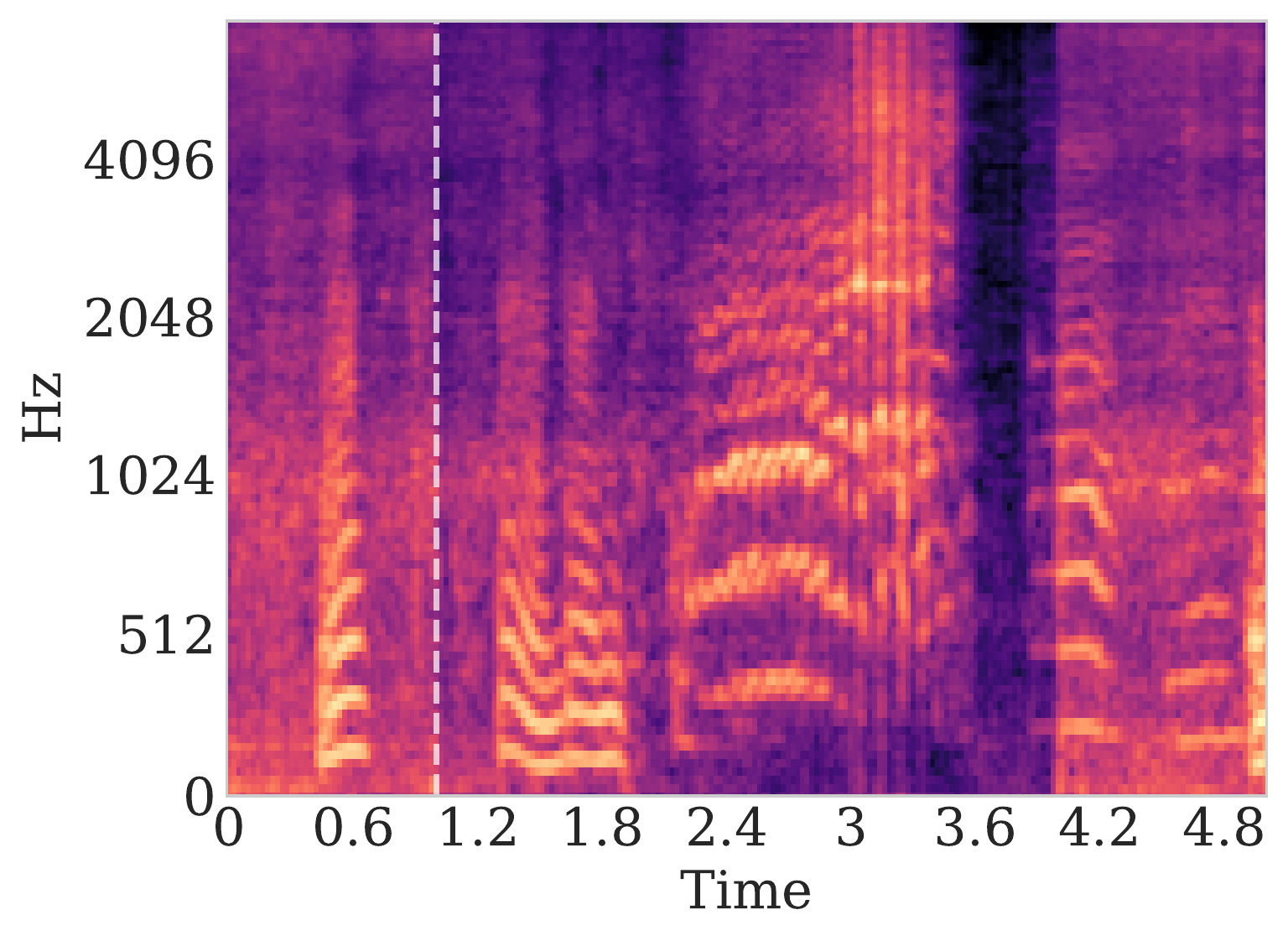}
    \label{fig:second}
\end{subfigure}
\caption{\label{fig:audio_cont_ex} A visual example of the text guided audio continuation. We plot Mel-Spectrograms of both audio continuation without text conditioning (left), and textually guided audio continuation (right). Input text is: ``speech and a goat bleating''.}
\end{figure}

We present a visual qualitative depiction of conditional and unconditional audio continuation in Figure~\ref{fig:audio_cont_ex}. We use the following text as conditioning, ``speech and a goat bleating'', and the first second of its corresponding audio as prompt. We mark the start of the generated segment using a dashed white line. In Figure~\ref{fig:audio_cont_ex} (left) we visualize unconditional audio continuation (i.e., no text). In Figure~\ref{fig:audio_cont_ex} (right) we visualize audio continuation conditioned text. As the audio prompts contains only human speech, the unconditional model generates a continuation that contains speech utterances, but does not generate any goat sounds. In contrast, the model that is conditioned on text successfully generates both human speech and goat sounds (left).

In table~\ref{tab:full_audio_cont} we report the complete results presented in Figure~\ref{fig:audio_cont}.
\begin{table}[h!]
    \centering
    \resizebox{0.8\columnwidth}{!}{%
    \begin{tabular}{l|c|c|c|c}
        \toprule
        \textsc{Text} & \textsc{Audio Prompt duration (sec.)} & \textsc{FAD}  & \textsc{KL} w. \rnd & KL w. \src \\
        \toprule
        \src & 0.5 & 1.97 & - & 1.49 \\
        no text	& 0.5 &	5.66 & 5.04 & 3.39 \\
        \rnd & 0.5 & 3.08 & 2.38 & 3.85 \\
        \midrule
        \src & 1 & 1.96 & - & 1.35 \\
        no text	& 1 & 4.70 & 4.95 & 2.86 \\
        \rnd & 1 & 3.17 & 2.65 & 3.25 \\
        \midrule
        \src & 2 & 1.92 & - & 1.21 \\
        no text & 2 & 3.15 & 5.00 & 2.10 \\
        \rnd & 2 & 2.92 & 3.13 & 2.47 \\
        \midrule
        \src & 3 & 1.90 & - & 1.09 \\
        no text & 3 & 2.46 & 5.04 & 1.58 \\
        \rnd & 3 & 2.58 & 3.54 & 1.84 \\
        \midrule
        \src & 4 & 1.97 & - & 1.00 \\
        no text & 4 & 2.14 & 5.10 & 1.18 \\
        \rnd & 4 & 2.29 & 4.37 & 1.34 \\
        \bottomrule
    \end{tabular}
    }
    \caption{Full results of the FAD and KL metrics for all text condition and audio prompt settings}
    \label{tab:full_audio_cont}
\end{table}

Finally, we analyze the effect of model sizes when considering text encoder and \ac{ALM}, on the generated audio. In Table~\ref{tab:model_size} we report KL and FAD scores for four different combinations: $\{\textrm{T5-base, T5-large}\} \times \{\textrm{ALM-base, ALM-large}\}$. When using a larger T5 encoder we observe a big improvement in terms KL, and minor improvement in FAD. On the other hand, when using larger ALM, we see a similar improvement in terms or KL with a significantly bigger improvement in FAD. Using both T5-large and ALM-large yields the best results overall.

\begin{table}[h!]
    \centering
    \begin{tabular}{l|l|c|c}
        \toprule
        \textsc{T5}    & \textsc{ALM}   & \textsc{KL} & \textsc{FAD} \\ 
        \toprule
        Base  & Base  &  2.09  & 3.13 \\ 
        Base  & Large &  1.92  & 2.27 \\ 
        Large & Base  &  1.91  & 3.03 \\ 
        Large & Large &  1.69  & 1.82 \\ 
        \bottomrule
    \end{tabular}
    \caption{Ablation study. We report KL and FAD scores for four different text-encoder and ALM setups, considering base and large models.}
    \label{tab:model_size}
\end{table}
\end{document}